\documentclass[draft]{agujournal2019}
\usepackage{url} 
\usepackage{lineno}
\usepackage[inline]{trackchanges} 
\usepackage{soul}
\usepackage{multirow}
\usepackage{amsmath,bm}
\usepackage{amssymb}
\usepackage{xr}
\usepackage{tabularx}
\usepackage[flushleft]{threeparttable}

\draftfalse

\journalname{AGU Journal}

\begin{document}
\justify

\title{Improving prediction of the terrestrial water storage anomalies during the GRACE and GRACE-FO gap with Bayesian convolutional neural networks}

\authors{Shaoxing Mo\affil{1}, Yulong Zhong\affil{2}, Xiaoqing Shi\affil{1}, Wei Feng\affil{3,4}, Xin Yin\affil{5}, and Jichun Wu\affil{1}}

\affiliation{1}{Key Laboratory of Surficial Geochemistry of Ministry of Education, School of Earth Sciences and Engineering, Nanjing University, Nanjing, China}
\affiliation{2}{School of Geography and Information Engineering, China University of Geosciences (Wuhan), Wuhan, China}
\affiliation{3}{State Key Laboratory of Geodesy and Earth’s Dynamics, Institute of Geodesy and Geophysics, Innovation Academy for Precision Measurement Science and Technology, Chinese Academy of Sciences, Wuhan, China}
\affiliation{4}{School of Geospatial Engineering and Science, Sun Yat-Sen University, Zhuhai, China}
\affiliation{5}{State Key Laboratory of Hydrology-Water Resources and Hydraulic Engineering, Nanjing Hydraulic Research Institute, Nanjing, China}

\correspondingauthor{Xiaoqing Shi}{shixq@nju.edu.cn}
\correspondingauthor{Wei Feng}{fengwei@whigg.ac.cn}

\begin{keypoints}
\item A Bayesian deep learning method is proposed for filling the gap between GRACE- and GRACE-FO-derived terrestrial water storage anomalies 
\item A state-of-the-art performance is obtained in bridging the gap at a global scale
\item Extreme dry and wet events during the gap are successfully recovered
\end{keypoints}

\begin{abstract}
The Gravity Recovery and Climate Experiment (GRACE) satellite and its successor GRACE Follow-On (GRACE-FO) provide valuable and accurate observations of terrestrial water storage anomalies (TWSAs) at a global scale. However, there is an approximately one-year observation gap of TWSAs between GRACE and GRACE-FO. This poses a challenge for practical applications, as discontinuity in the TWSA observations may introduce significant biases and uncertainties in the hydrological model predictions and consequently mislead decision making. To tackle this challenge, a Bayesian convolutional neural network (BCNN) driven by climatic data is proposed in this study to bridge this gap at a global scale. Enhanced by integrating recent advances in deep learning, including the attention mechanisms and the residual and dense connections, BCNN can automatically and efficiently extract important features for TWSA predictions from multi-source input data. The predicted TWSAs are compared to the hydrological model outputs and three recent TWSA prediction products. The comparison suggests the superior performance of BCNN in providing improved predictions of TWSAs during the gap in particular in the relatively arid regions. The BCNN's ability to identify the extreme dry and wet events during the gap period is further discussed and comprehensively demonstrated by comparing with the precipitation anomalies, drought index, ground/surface water levels. Results indicate that BCNN is capable of offering a reliable solution to maintain the TWSA data continuity and quantify the impacts of climate extremes during the gap. 
\end{abstract}

\section{Introduction}\label{sec:introduction}
The Gravity Recovery and Climate Experiment (GRACE) satellite and its successor GRACE Follow-On (GRACE-FO) provide unprecedentedly accurate observations of the spatio-temporal dynamics of terrestrial water storage anomalies (TWSAs) at a global scale. These TWSA observations have been widely utilized together with hydrological models to assess water cycle, droughts and floods, the impacts of changing climate on terrestrial water storage, etc~\cite{AghaKouchak2015,Fengrs10050674,Gentine_2019,Rateb2020,Richey2015,Rodell2018Natur,SOLTANI2021,Tapley2019}. Particularly, their applications in groundwater-related studies (e.g., the groundwater depletion and land subsidence problems in North China Plain and California's Central Valley) are more extensive due to the difficult-to-observe nature of groundwater systems, substantially augmenting our knowledge toward the systems and thus benefiting groundwater resources management~\cite<e.g.,>{Chang2020wrr,CHEN2014130,Famiglietti2011,Feng2013WRR,Fengrs10050674,Lietal2019WRR,Smith2020,Zhong_2018}. 

However, there is an approximately one-year gap of TWSA observations between the GRACE and GRACE-FO missions. Considering that the TWSA observations are usually assimilated in the hydrological models for higher reliability~\cite{Lietal2019WRR,NieWRR2019,SOLTANI2021,YIN2020125348,Zaitchik}, discontinuity in the time series observations may introduce significant biases and uncertainties in the model predictions and consequently mislead decision making~\cite{AlexSun2020}. This is especially the case when there exist extreme dry or wet events during the gap as they usually cause abnormal changes in the TWSA signals. Bridging this gap is thus of crucial importance for practical applications. There have been many efforts undertaken to predict/reconstruct GRACE TWSAs at regional or global scales using data-driven methods~\cite <e.g.,>[]{Ahmed_2019RS,Forootan2014,Forootanrs12101639,Humphrey2017GRL,Humphrey-2019-REC,Jing2020,LietalWRR2020,LONG2014145,Richter2021,AlexSun2019,AlexSun2020,SunZL2020WRR,WANG2021125972}. While these studies have generally obtained relatively good performances in humid regions, the performance in the relatively arid regions remains to be improved (a global aridity index map and the spatial distribution of hyper-arid, arid, semi-arid, semi-humid, and humid regions are shown in Figure~\ref{fig:AI}), calling for innovative solutions. 

The objectives of this work are to provide improved predictions of TWSAs to fill the gap between GRACE and GRACE-FO at a global scale (excluding Antarctica), so as to maintain the TWSA observation continuity and subsequently benefit the related hydrological applications. The major challenges associated with this task are twofold. The first one is the difficulty in modeling the long-term TWSA trends caused by anthropogenic activities, which accounts for the decreased performance of existing methods in the relatively arid regions~\cite{Humphrey-2019-REC,LietalWRR2020,SunZL2020WRR}. Since multiple predictors are usually used in the prediction models, another challenge is how to emphasize important predictors and suppress unnecessary ones, so that the models can effectively extract informative features for TWSA predictions~\cite{LietalWRR2020,AlexSun2020}.

In this study, the first challenge is tackled by making use of the long-term trends retrieved from the available GRACE/GRACE-FO data. Recent years have witnessed a rapid development of deep learning and its impressive performance in a variety of applications~\cite{GU2018354,Reichstein2019Natur,Shen2018}. The second challenge is solved by developing a deep learning-based prediction model. To this end, we propose a Bayesian convolutional neural network (BCNN) driven by climatic inputs to bridge the GRACE and GRACE-FO gap. The convolutional neural networks (CNNs) have exhibited an outstanding capability in extract underlying features from images and learning complex relationships between the image data~\cite{GU2018354,Mo2019_co2,Mo2019inverse,Mo2020,Shen2018,AlexSun2019}. In BCNN, the input climatic data and the target GRACE TWSAs are treated as images, enabling the model to exploit and utilize the spatially correlated features in the images for TWSA predictions and handle simultaneously the global scale. \citeA{AlexSun2019} applied CNNs for prediction of TWSAs in India and the CNN outperformed the hydrological models in providing more accurate TWSA estimates. The development of our BCNN leverages and combines the recent advances in deep learning, including the channel and spatial attention mechanisms~\cite{Woo_2018_ECCV}, residual~\cite{He_2016_CVPR} and dense~\cite{Huang_2017_CVPR} connection modules, Bayesian training strategy~\cite{LiuSVGD2016,ZHU2018415} and other~\cite{GU2018354}. By comparing with the hydorlogical model outputs and three recent TWSA prediction products, we will show that the combination of these strategies enables BCNN to automatically and efficiently extract informative features from multi-source driving data, offer predictive uncertainties, and consequently achieve a state-of-the-art performance in filling the gap. The BCNN-predicted TWSAs are then used to identify the extreme dry/wet events during the gap and the identified events are demonstrated using the precipitation anomalies, drought index, ground/surface water levels. 

The rest of the paper is organized as follows. The data used are described in section~\ref{sec:data}. In section~\ref{sec:methods}, the BCNN model, including its architecture design and training data, is introduced. In section~\ref{sec:results&discussion}, we evaluate BCNN's performance by comparing with the hydrological model outputs and previous TWSA prediction products. The BCNN-identified extreme dry/wet events during the GRACE and GRACE-FO gap are also analyzed and demonstrated in this section. Finally, the conclusions are summarized in section~\ref{sec:conclusions}.

\section{Data and Processing}\label{sec:data}

\subsection{GRACE TWSA Data}\label{sec:GRACE}
The monthly GRACE mascon product released by the Jet Propulsion Laboratory (JPL) (available at \url{https://podaac.jpl.nasa.gov/dataset/TELLUS_GRAC-GRFO_MASCON_CRI_GRID_RL06_V2}), which has a spatial resolution of $0.5^{\circ}\times0.5^{\circ}$~\cite{Watkins2015}, is used in this study. The JPL GRACE TWSA data are provided as anomalies with respect to the 2004 to 2009 average. The observations cover two periods, that is, April 2002-June 2017 (i.e. the GRACE mission) and June 2018-present (i.e. the GRACE-FO mission), with a 11-month gap in between. In addition, there are some intermittent one or two months with missing TWSA data within each mission, these missing months are interpolated using the data of neighboring months. Our aim is to fill the 11-month gap (i.e. July 2017-May 2018) for the land areas with the BCNN method. To facilitate the comparison with previous GRACE prediction studies, we resampled averagely the data to $1^{\circ}\times1^{\circ}$ grid. 

\subsection{ERA5-Land Driving Data}\label{sec:driving_data}
The driving data used to predict the GRACE TWSAs are extracted from the European Centre for Medium-Range Weather Forecasts (ECMWF) ERA5-Land (ERA5L) dataset (available at \url{https://cds.climate.copernicus.eu}) \cite{ERA5L2019}. The ERA5L dataset contains an improved version of the land components of the ERA5 climate reanalysis, making it more accurate for land applications. The data are provided at a spatial resolution of $0.1^{\circ}\times0.1^{\circ}$ and cover a period from January 1981 to near present. The driving data contain four predictor variables, including the monthly precipitation, temperature, cumulative water storage changes (CWSCs), and ERA5L-derived TWSAs. The spatial resolution of these data is averagely resampled into to $1^\circ\times1^\circ$ to be consistent with GRACE TWSAs.

The water storage change (WSC) is calculated as the difference between the inflow (i.e. precipitation $P$) and outflows (i.e., evapotranspiration $ET$ and runoff $RO$) of a grid cell based on the water balance:
\begin{linenomath*}
\begin{equation}
    \mathrm{WSC} = P - ET - RO.
\end{equation}
\end{linenomath*}
The CWSC is thus written as follows:
\begin{linenomath*}
\begin{equation}
    \mathrm{CWSC}_t =\sum_{i=1}^t \mathrm{WSC}_i=\sum_{i=1}^t(P_i - ET_i - RO_i),
\end{equation}
\end{linenomath*}
where $t$ denotes the month index. CWSC correlates well to GRACE TWSA in most regions as shown in Figure~\ref{fig:R_CWSCvsGRACE}. Therefore, it is used as an additional predictor of GRACE TWSAs. 

The ERA5L dataset includes water storage in soil moisture (in four layers spanning from 0 to 289~cm), snow, and canopy. Thus, the ERA5L TWSAs are calculated by summing these components and then subtracting the long-term mean between 2004 and 2009 to be consistent with GRACE TWSAs, as represented by:
\begin{linenomath*}
\begin{equation}
    \mathrm{TWSA_{\rm{ERA5L}} = SMS+SWS+CWS-\overline{TWS}_{0409}},
\end{equation}
\end{linenomath*}
where SMS, SWS, and CWS are soil moisture storage, snow water storage, and canopy water storage, respectively, $\overline{\rm{TWS}}_{0409}$ denotes 2004-2009 mean of the three components.

\subsection{Time Series Data detrending}\label{sec:detrend}
The GRACE TWSA time series in the relatively arid regions often exhibits long-term declining/rising trends caused by the human intervention  and/or changing climate. This presents challenges for the TWSA prediction tasks, as the driving data (e.g., the climatic data or hydrological model outputs) may not be able to well reflect the influences of these factors~\cite{Humphrey-2019-REC,LietalWRR2020,SunZL2020WRR}. For the 11-month gap (July 2017-May 2018) filling task considered here, fortunately, the GRACE TWSA data before (April 2002-June 2017) and after (June 2018-) the gap are available. Thus, we can obtain directly the long-term trend signals for the missing interval from existing data. Then we predict in the gap filling task the detrended TWSA signals instead, which are generally less challenging relative to predicting the original signals~\cite{Humphrey-2019-REC,LietalWRR2020}.  Mathematically, the GRACE TWSAs are decomposed via linear detrending into two components:
\begin{linenomath*}
\begin{equation}
\label{eq:TWSA_detrend}
    \mathrm{TWSA_{GRACE}=TWSA_{GRACE}^{detrend}+trend_{GRACE}},
\end{equation}
\end{linenomath*}
where $\mathrm{TWSA_{GRACE}^{detrend}}$ and $\mathrm{trend_{GRACE}}$ are the detrended data and a linear trend term, respectively. Correspondingly, the driving data described in section~\ref{sec:driving_data} are also detrended. In the prediction task, the BCNN model learns to predict the $\mathrm{TWSA_{GRACE}^{detrend}}$ signals. Finally, the predictions for the original TWSAs are obtained by adding the GRACE trend:
\begin{linenomath*}
\begin{equation}
\label{eq:prediction_add_trend}
    \mathrm{TWSA_{BCNN}=TWSA_{BCNN}^{detrend}+trend_{GRACE}}.
\end{equation}
\end{linenomath*}

\section{Methods}\label{sec:methods}
\subsection{BCNN Deep Learning Models}\label{sec:BCNN}
We propose a BCNN method to learn the underlying relationship between $\mathrm{TWSA_{GRACE}^{detrend}}$ and its four predictors (i.e., the detrended $P$, $T$, $\mathrm{CWSC}$, and TWSA$_{\mathrm{ERA5L}}$). Without loss of generality, here we use $\mathbf{x}$ and $\mathbf{y}$ to denote the network inputs (i.e. predictors) and outputs (i.e. $\mathrm{TWSA_{GRACE}^{detrend}}$), respectively. In BCNN, the inputs and outputs are both organized as images (matrices) and the learning task becomes an image regression problem:
\begin{linenomath*}
\begin{equation}\label{eq:image-regression}
    \bm{\eta}:~\mathbf{x}\in\mathbb{R}^{n_{x}\times H\times W}\longrightarrow\mathbf{y}\in\mathbb{R}^{n_{y}\times H\times W},
\end{equation}
\end{linenomath*}
where $\bm{\eta}=\bm{\eta}(\mathbf{x},\mathbf{w})$ is a BCNN model, with $\mathbf{w}$ denoting all trainable network parameters, including the weights and biases. The inputs $\mathbf{x}$ and outputs $\mathbf{y}$ become $n_x$ and $n_y$ images, respectively, all with $H\times W$ pixels (grids). 

The network predictions are inevitably associated with epistemic uncertainties induced by lack of training data. Failing to estimate the predictive uncertainty may lead to overconfident results. This is especially the case for the GRACE TWSA prediction task considered here, as the available training data are limited. Contrary to vanilla CNNs which treat the network parameters $\mathbf{w}$ as deterministic unknowns and thus fail to offer the predictive uncertainties, in BCNN $\mathbf{w}$ are treated as random variables. Given a set of training data $\mathcal{D}=\left\{\mathbf{x}_i,\mathbf{y}_i\right\}_{i=1}^{N_{\rm{train}}}$, the network training is to infer the posterior distribution of $\mathbf{w}$, $p(\mathbf{w}|\mathcal{D})$. Consequently, one can obtain the prediction distribution of the target variables $\mathbf{y}$: $p(\mathbf{y}|\mathbf{w})$, $\mathbf{w}\sim p(\mathbf{w}|\mathcal{D})$, and in particular the mean $\mathbb{E}(\mathbf{y}|\mathbf{w})$ and standard deviation $\rm{Std}(\mathbf{y}|\mathbf{w})$. 

In BCNN, the Bayesian training strategy proposed in~\citeA{ZHU2018415} is employed. Mathematically, the BCNN model is expressed as follows:
\begin{linenomath*}
\begin{equation}\label{eq:cnn}
    \hat{\mathbf{y}}=\bm{\eta}(\mathbf{x},\mathbf{w})+\mathbf{n}(\mathbf{x},\mathbf{w}),
\end{equation}
\end{linenomath*}
where $\hat{\mathbf{y}}$ denotes the BCNN predictions and $\mathbf{n}(\cdot)$ is an additive Gaussian noise term modeling the aleatoric uncertainty. A variational Bayesian inference algorithm called stein variational gradient descent (SVGD)~\cite{LiuSVGD2016}, which is similar to standard gradient descent while maintaining the particle methods' high efficiency, is utilized to estimate the posterior distribution $p(\mathbf{w}|\mathcal{D})$. In implementation, we use $N_S$ samples of $\mathbf{w}$ to approximate the posterior distribution $p(\mathbf{w}|\mathcal{D})$. The $N_S$ samples $\left\{\mathbf{w}_i\right\}_{i=1}^{N_S}$ are respectively optimized using the Adam optimizer~\cite{KingmaB14}, whose gradient derives from SVGD. The predictive mean and standard deviation (i.e. uncertainty) of BCNN for an arbitrary input $\mathbf{x}$ are then calculated based on the $N_S$ predictions ($\hat{\mathbf{y}}^{(i)}=\bm{\eta}(\mathbf{x},\mathbf{w}_i)+\mathbf{n}(\mathbf{x},\mathbf{w}_i),\:i=1,\ldots,N_S$). For more details regarding the SVGD Bayesian training strategy, one can refer to~\citeA{LiuSVGD2016} and ~\citeA{ZHU2018415}.

\subsection{BCNN Architecture Design and Training}\label{sec:net_arc_train}
There are two major challenges associated with designing an effective deep learning model for predicting the GRACE TWSAs. The first one is to effectively extract useful features and suppress unnecessary ones for TWSA predictions from multiple predictors. The second one is to build a reliable prediction model given limited GRACE training data (so far only $\sim$200 months of GRACE data are available). In BCNN, we tackle the two challenges by leveraging the recent advances in deep learning, as described below.

\begin{figure}[h!]
    \centering
    \includegraphics[width=\textwidth]{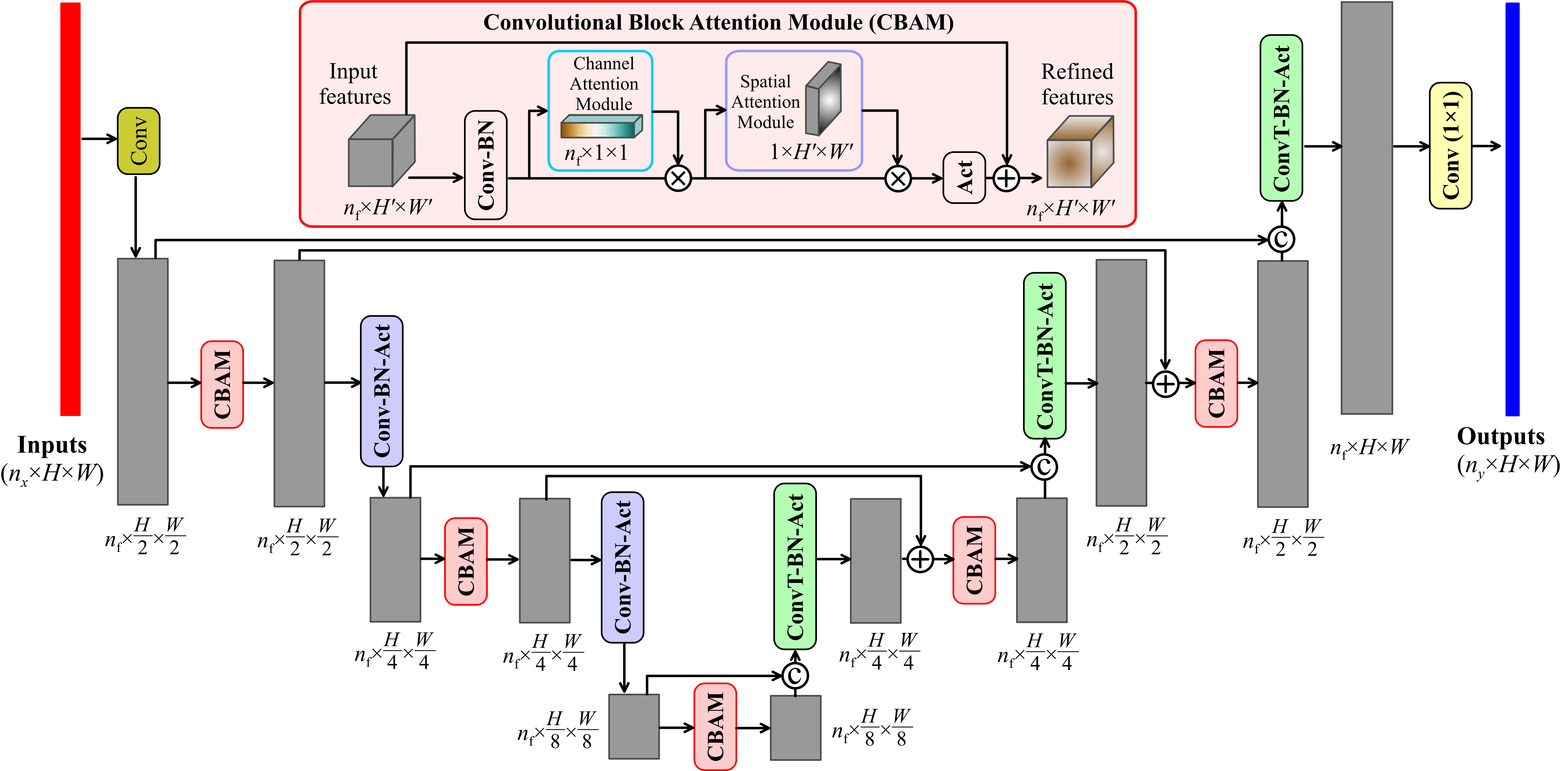}
    \caption{Illustration of the Bayesian convolutional neural network (BCNN) architecture. BCNN takes $n_{x}$ images with a size of $H\times W$ as inputs and generates $n_{y}$ images with the same size. It is an alternating  cascade of convolutional (Conv)/transposed convolutional (ConvT) layers and convolutional block attention modules (CBAM), each of which outputs $n_{\rm{f}}=48$ feature maps. The size of feature maps is sequentially halved in each Conv layer from $H\times W$ to $\frac{H}{8}\times\frac{W}{8}$ so as to extract multi-scale features, and then sequentially recovered to $H\times W$ using the ConvT layers. The symbols \textcircled{$+$}, \textcircled{c}, and \textcircled{$\times$} denote the addition (i.e. residual connection), concatenation (i.e. dense connection), and multiplication (i.e. attention connection) operations, respectively. Act=activation; BN=batch normalization.}
    \label{fig:BCNN}
\end{figure}

The BCNN network architecture is depicted in Figure~\ref{fig:BCNN}. The convolutional block attention module (CBAM) proposed in~\citeA{Woo_2018_ECCV} is used as the basic block. Given $n_x$ images with a resolution of $H\times W$ as inputs to the network, they are passed through an alternating cascade of convolutional/transposed convolutional layers and CBAMs, each of which produces $n_{\mathrm{f}}$ $H^{'}\times W^{'}$ feature maps, to extract multi-scale features to finally predict $n_y$ images for the targets. The CBAM block contains two attention modules, namely the channel and spatial attention modules as depicted in Figure~\ref{fig:BCNN} and detailed in Figures~\ref{fig:channel_spatial_attention}. More specifically, the channel module outputs $n_{\rm{f}}$ weights between 0 and 1 assigning to the $n_{\rm{f}}$ feature maps to tell the network `what' (i.e. which maps) to attend; the spatial module outputs a $H^{'}\times W^{'}$ weight matrix assigning to the $H^{'}\times W^{'}$ pixel feature maps to tell the network `where'  (i.e. which regions) to emphasize or suppress. As such, the network is able to automatically focus on important features and suppress unnecessary ones~\cite{Woo_2018_ECCV}. 

The residual connection~\cite{He_2016_CVPR} and dense connection~\cite{Huang_2017_CVPR} strategies have been shown to be effective in enhancing information flow through the network and thus substantially improving the performance. Therefore, they are implemented in our BCNN model. In particular, in the residual connection structure, the feature maps with the same shape ($n_{\mathrm{f}}\times H^{'}\times W^{'}$) but at different layers are connected by applying element-wise addition~\cite{He_2016_CVPR}. In the dense connection structure, the feature maps with the same size ($H^{'}\times W^{'}$) but at different layers are cascaded together and subsequently fed as inputs into the next layer~\cite{Huang_2017_CVPR} (Figure~\ref{fig:BCNN}). The Mish function~\cite{Misra2019Mish} is employed in BCNN as the activation function of hidden layers unless otherwise stated.

We use twelve years of monthly GRACE TWSA data from April 2002 to March 2014 (i.e. 144 months, $\sim$69$\%$) to train the BCNN network, and those from April 2014 to June 2017 and June 2018 to August 2020 (i.e. 66 months, $\sim$31$\%$) to test the performance. We set the number of lags for predictors to 2. That is, for month $t$, the inputs to BCNN are the detrended $P_i$, $T_i$, CWSC$_i$, and TWSA$_{\mathrm{ERA5L},i}$ with $i=t-2,...,t$. Thus, each sample contains $n_{x}=12$ input images and $n_{y}=1$ output image (i.e. TWSA$_{\mathrm{GRACE},t}^{\mathrm{detrend}}$). The region spanning from $60^{\circ}$S to $84^{\circ}$N and $180^{\circ}$W to $180^{\circ}$E represented by a $H\times W=144\times360$ image is considered. For non-land pixels (grids), the pixel values are set to a constant of 0. During network training, we use $N_S=20$ samples of $\mathbf{w}$ in the SVGD algorithm to approximate the posterior distribution, as suggested in~\citeA{ZHU2018415}. The network is trained for 200 epochs, with a mean squared error loss function quantifying the predictive accuracy, an initial learning rate of 0.0025 in the Adam optimizer and a batch size of 12. The training is performed on a single GPU (NVIDIA Tesla V100) and takes $\sim$80~minutes. The network performance is evaluated based on the test dataset using three commonly used metrics, including the correlation coefficient ($R$), Nash-Sutcliffe efficiency coefficient (NSE), and normalized root mean squared error (NRMSE), which are defined as follows:
\begin{linenomath*}
\begin{equation}\label{eq:R}
    R=\frac{\sum_{i=1}^{N_{\mathrm{test}}}\left(y_{i}-\bar{y}\right)\left(\hat{y}_{i}-\bar{\hat{y}}\right)}{\sqrt{\sum_{i=1}^{N_{\mathrm{test}}}\left(y_{i}-\bar{y}\right)^{2}} \sqrt{\sum_{i=1}^{N_{\mathrm{test}}}\left(\hat{y}_{i}-\bar{\hat{y}}\right)^{2}}},
\end{equation}
\end{linenomath*}

\begin{linenomath*}
\begin{equation}\label{eq:NSE}
    \mathrm{NSE}=1-\frac{\sum_{i=1}^{N_{\mathrm{test}}}\left(y_i-\hat{y}_i\right)^{2}}{\sum_{i=1}^{N_{\mathrm{test}}}\left(y_i-\bar{y}\right)^{2}},
\end{equation}
\end{linenomath*}

\begin{linenomath*}
\begin{equation}\label{eq:RMSE}
    \mathrm{NRMSE}=\frac{\sqrt{\frac{1}{N_{\mathrm{test}}} \sum_{i=1}^{N_{\mathrm{test}}}\left(y_{i}-\hat{y}_{i}\right)^{2}}}{y_{\max}-y_{\min}},
\end{equation}
\end{linenomath*}
where $y$ denotes the observations, with a mean value of $\bar{y}$, $\hat{y}$ denotes the predictions, with a mean value of $\bar{\hat{y}}$, $N_{\mathrm{test}}$ is the number of test samples, $y_{\max}$ and $y_{\min}$ represent the maximum and minimum values of the observations, respectively. The $R$ or NSE values closer to 1.0 or lower NRMSE values indicate better performances.

\section{Results and Discussion}\label{sec:results&discussion}

\subsection{Accuracy Assessment}\label{sec:accuracy_assess}
To illustrate the performance of BCNN, the $R$, NSE, and NRMSE metrics are also computed for the ERA5L TWSAs and the hydrological model Noah-derived TWSAs ($1^{\circ}\times 1^{\circ}$, version 2.1)~\cite{Noah}.

\begin{figure}[h!]
    \centering
    \includegraphics[width=\textwidth]{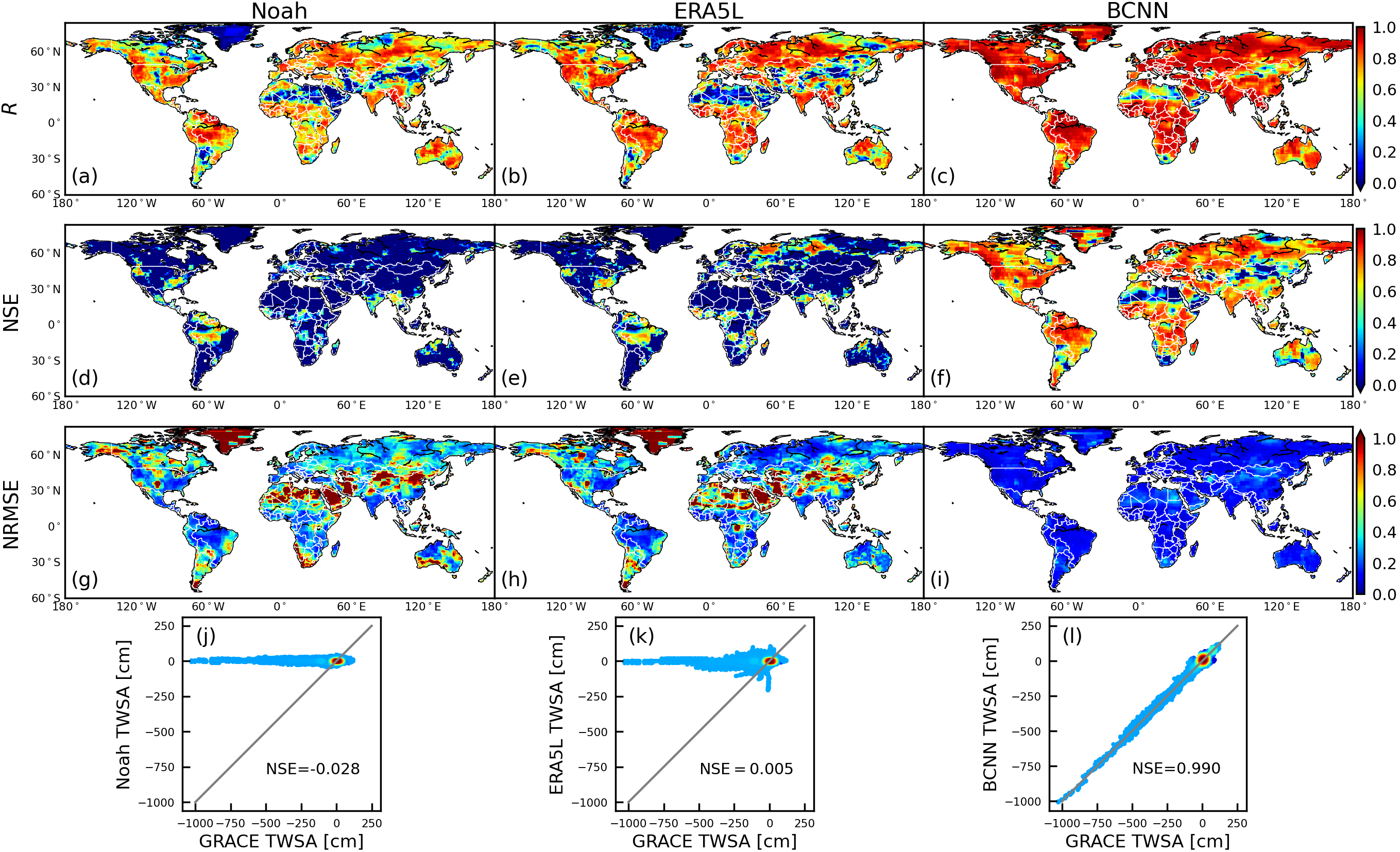}
    \caption{Spatial maps of $R$ (row 1), NSE (row 2), and NRMSE (row 3) values between the GRACE TWSAs and the Noah- (column 1), ERA5L- (column 2), and BCNN-derived (column 3) TWSAs during the test period (April 2014-June 2017, June 2018-August 2020). (j-l) The density scatter plots between the GRACE and modeled TWSAs.}
    \label{fig:R_NSE_RMSE}
\end{figure}

Figure~\ref{fig:R_NSE_RMSE} shows the spatial distributions of the accuracy metrics obtained by Noah, ERA5L, and BCNN. While Noah and ERA5L TWSAs both show relatively good correlations with GRACE TWSAs in most regions except Greenland and the hyper-arid areas like Sahara, Gobi, and Arabian (Figures~\ref{fig:R_NSE_RMSE}a and~\ref{fig:R_NSE_RMSE}b), BCNN TWSAs exhibit clearly better correlations with GRACE TWSAs in almost all regions (Figure~\ref{fig:R_NSE_RMSE}c). For the NSE metric, which measures directly the matching quality between the predicted and observed values, both Noah and ERA5L obtain relatively low values ($<$0) in most regions except in some humid regions like Amazon and Southeastern United States (Figures~\ref{fig:R_NSE_RMSE}d and~\ref{fig:R_NSE_RMSE}e). In contrast, BCNN provides relatively high values ($>$0.5) in most regions (Figure~\ref{fig:R_NSE_RMSE}f). Note that while BCNN provides a higher accuracy than Noah and ERA5L in the hyper-arid regions, the NSE values are still low relative to other regions. This is due to the fact that the variability of TWSAs in these regions is dominated by noise~\cite{Humphrey2016}. The improved performance of BCNN over Noah and ERA5L can be also illustrated by the NRMSE maps (Figure~\ref{fig:R_NSE_RMSE}(g-i)) and the density scatter plots between the modeled and GRACE TWSAs (Figure~\ref{fig:R_NSE_RMSE}(j-l)), which indicate that BCNN reduces significantly the NRMSE errors and its predicted TWSAs display a good consistency with GRACE TWSAs with an overall NSE value of 0.990, much higher than those of Noah (-0.028) and ERA5L (0.005).

\begin{figure}[h!]
    \centering
    \noindent\includegraphics[width=\textwidth]{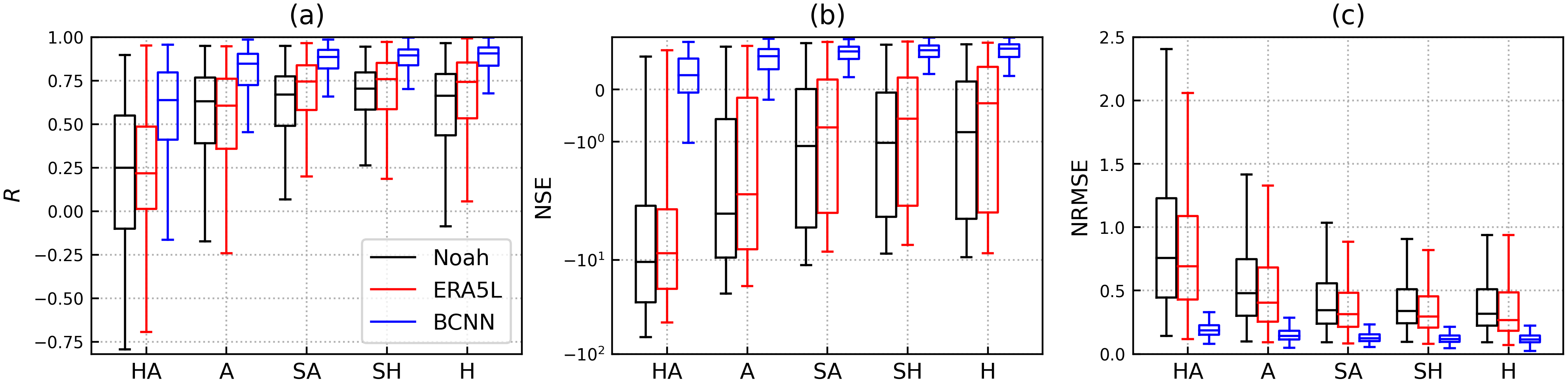}
    \caption{Boxplots of the (a) $R$, (b) NSE, and (c) NRMSE values at the grid cells in the hyper-arid (HA), arid (A), semi-arid (SA), semi-humid (SH), and humid (H) regions obtained by Noah, ERA5L, and our BCNN. The outliers are not shown.}
    \label{fig:boxplot_vsNoah_ERA5L}
\end{figure}

It can be found from Figure~\ref{fig:R_NSE_RMSE} that the performance is highly dependent on the regional climatic conditions. We further compare the three models' $R$, NSE, and NRMSE values at the grid cells in the hyper-arid, arid, semi-arid, semi-humid, and humid regions (the spatial distribution of these regions is shown in Figure~\ref{fig:AI}b). The results are summarized in the boxplots depicted in Figure~\ref{fig:boxplot_vsNoah_ERA5L}, with their medians being listed in Table~\ref{tab:medians_BCNNvsNoah&ERA5L}. In general, higher performances are found as expected in the regions with more humid climate. This is probably because the relatively arid regions usually have low signal-to-noise ratios but are often associated with heavy human interventions (e.g., groundwater extractions and reservoir operations) \cite{Humphrey2016,SunZL2020WRR}. Again, BCNN clearly outperforms Noah and ERA5L in providing much higher accuracy in all five classes of regions. For example, the median NSE  values of Noah, ERA5L, and BCNN in the hyper-arid region are -10.501, -8.509, and 0.273, respectively (Table~\ref{tab:medians_BCNNvsNoah&ERA5L}). 

The poor performance of Noah/ERA5L in predicting TWSAs is mainly due to their underestimation of the long-term trends of GRACE TWSAs~\cite{ScanlonE1080}. Considering the GRACE trends have been employed in BCNN (section~\ref{sec:detrend}), to illustrate that the outperformance of BCNN is not only obtained by simply utilizing the GRACE trend information, but more importantly attributes to its ability to discover informative features for TWSA predictions from multi-source data, we correct the Noah and ERA5L TWSAs with the GRACE trends. That is,
\begin{linenomath*}
\begin{equation}\label{eq:cNoah&cERA5L}
    \mathrm{TWSA_M}=\texttt{detrend}(\mathrm{TWSA_M})+\mathrm{trend_{GRACE}},
\end{equation}
\end{linenomath*}
where \texttt{detrend($\cdot$)} denotes the linear detrending operation, TWSA$_{\rm{M}}$ represents the Noah or ERA5L TWSAs. Figure~\ref{fig:metrics_Noah_ERA5L_trendGRACE} shows the same metrics as in Figure~\ref{fig:R_NSE_RMSE} and Figure~\ref{fig:boxplot_vsNoah_ERA5L_GRACEtrend} shows the similar boxplots as in Figure~\ref{fig:boxplot_vsNoah_ERA5L} for the corrected Noah and ERA5L TWSAs. The medians of the boxplots are summarized in Table~\ref{tab:medians_BCNNvsNoah&ERA5L}. The comparison between Figures~\ref{fig:R_NSE_RMSE}(c,f,i,l) and~\ref{fig:metrics_Noah_ERA5L_trendGRACE} and the results presented in Figure~\ref{fig:boxplot_vsNoah_ERA5L_GRACEtrend} and Table~\ref{tab:medians_BCNNvsNoah&ERA5L} again suggest BCNN's higher performance, although the consistency between the corrected Noah/ERA5L TWSAs and GRACE TWSAs has been improved significantly as expected.

\begin{figure}[h!]
    \centering
    \includegraphics[width=\textwidth]{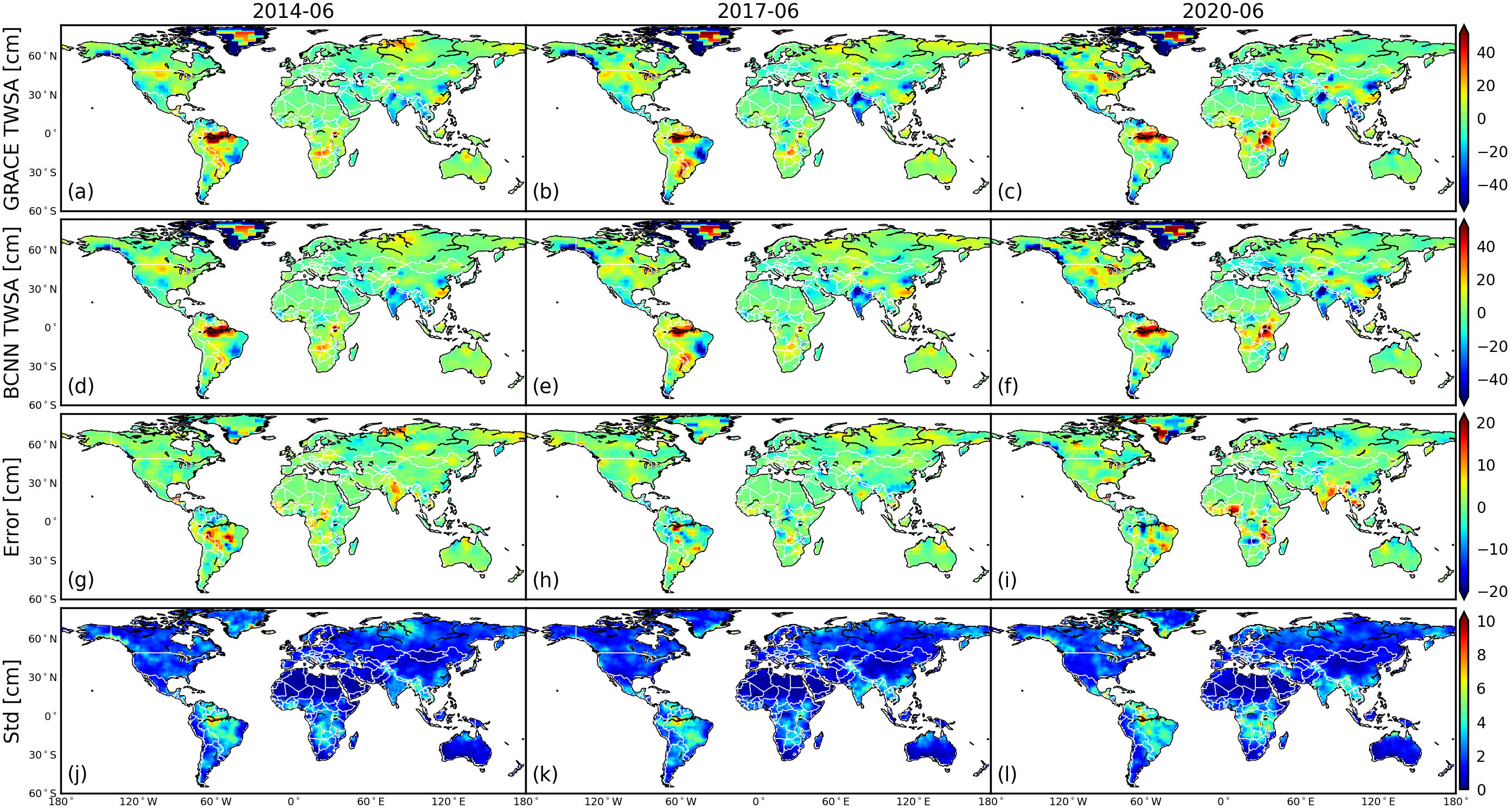}
    \caption{The GRACE (a-c) and BCNN (d-f) TWSAs in three test months (June 2014, June 2017, and June 2020). (g-i) The BCNN predicted error (i.e. TWSA$_{\mathrm{GRACE}}$-TWSA$_{\mathrm{BCNN}}$) and (j-l) standard deviation (Std). The Std is calculated based on an ensemble of $N_S=20$ BCNN predictions (section~\ref{sec:BCNN}). }
    \label{fig:CNN_pred}
\end{figure}

Figure~\ref{fig:CNN_pred} depicts BCNN's TWSA predictions for three test months in June 2014, June 2017, and June 2020. The three months cover the early-, mid-, and late-term of the test period, with June 2017 being the last month before the missing gap. For comparison, we also show the reference GRACE TWSAs. It can be seen that BCNN successfully captures the spatial patterns of GRACE TWSAs and provides close predictions in the three months (Figure~\ref{fig:CNN_pred}(a-f)). The predicted errors and uncertainties in humid regions (e.g., Amazon, Central Africa, South Asia, and Greenland) are generally larger compared to other regions (Figure~\ref{fig:CNN_pred}(g-l)), which are mainly because of the relatively high signal-to-noise ratio in the humid regions. The BCNN's TWSA predictions for all 66 test months, as well as the corresponding reference GRACE TWSAs, predicted errors and uncertainties, are shown in the GIF (Graphics Interchange Format) animation attached in the supporting information.

\begin{figure}[h!]
    \centering
    \includegraphics[width=\textwidth]{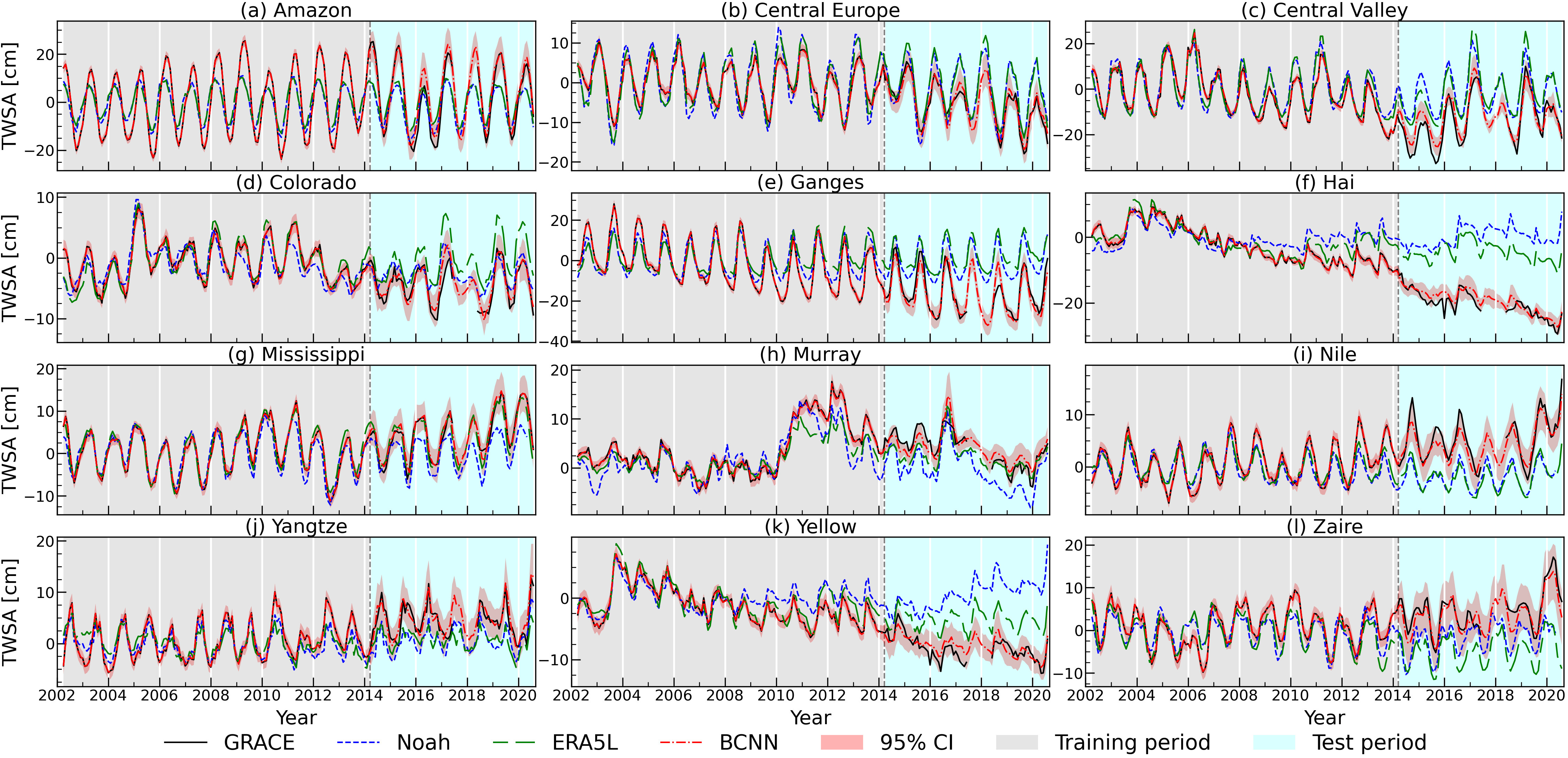}
    \caption{The region/basin-averaged GRACE, Noah, ERA5L and BCNN TWSA time series in different basins/regions. The red shaded area denotes the $95\%$ confidence interval (CI) of BCNN predictions.}
    \label{fig:pred_timeseries}
\end{figure}

The performance of BCNN in providing reliable predictions for TWSAs is further demonstrated in Figures~\ref{fig:pred_timeseries} and~\ref{fig:time_series_SI}, which depict the basin/region-averaged TWSA time series derived from GRACE, Noah, ERA5L, and BCNN in different basins/regions (Locations of these basins/regions are shown in Figure~\ref{fig:basin_map}). For completeness, we show the time series in both training and test periods. Again, the BCNN TWSAs fit obviously better with the reference GRACE TWSAs than Noah and ERA5L TWSAs.  While BCNN may slightly underestimate/overestimate some peak/valley values of GRACE TWSAs during the test period, the GRACE TWSA curves are almost completely enveloped within the $95\%$ prediction interval.

\subsection{Detection of the Extreme Dry and Wet Events during the Gap}\label{sec:drought_flood_detect}
The GRACE TWSA time series are effective indicators for detection of extreme dry/wet events, which cause unusual decreases/increases in the TWSA signals, and quantifying the water loss/gain during the events~\cite{Fengrs10050674,Humphrey2016,LietalWRR2020,Tapley2019}. For example, the GRACE successfully identified the 2016/2017 and 2018/2019 droughts in Central Europe (Figure~\ref{fig:pred_timeseries}b)~\cite{Boergens_etal_2020}, the droughts from 2012 to 2016 in Central Valley (Figure~\ref{fig:pred_timeseries}c)~\cite{Xiao_etal_2017}, and the flood in Summer 2020 in Yangtze River Basin (Figure~\ref{fig:pred_timeseries}j)~\cite{WEI2020100038}. As shown in Figure~\ref{fig:pred_timeseries}, the BCNN TWSAs agree well with GRACE TWSAs during these extreme events, suggesting that BCNN is capable of detecting the drought- and flood-induced abnormal TWSA signals. 

\begin{figure}[h!]
    \centering
    \includegraphics[width=\textwidth]{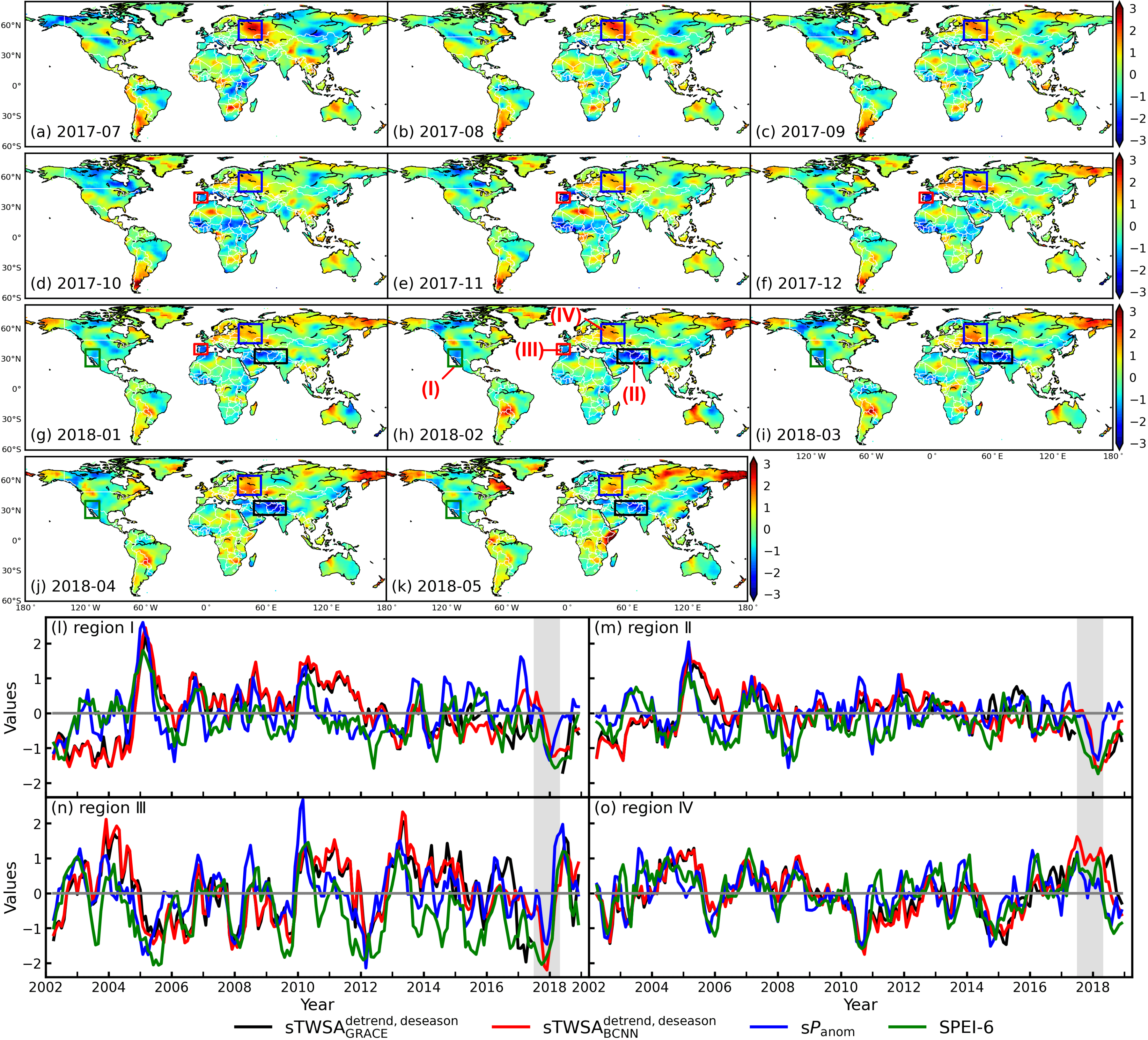}
    \caption{(a-k) The standardized signals of the detrended and deseasonalized BCNN TWSAs (sTWSA$_{\mathrm{BCNN}}^{\mathrm{detrend,deseason}}$) during the gap (July 2017-May 2018). The four regions within the rectangles (labeled with I-IV in h) have significant abnormal signals during this period. (l-o) The region-averaged time series of sTWSA$_{\mathrm{GRACE}}^{\mathrm{detrend,deseason}}$, sTWSA$_{\mathrm{BCNN}}^{\mathrm{detrend,deseason}}$, standardized precipitation anomalies (s$P_{\mathrm{anom}}$), and 6-month SPEI (SPEI-6) in the four regions. The shaded area denotes the gap period.}
    \label{fig:TWS_intersubseas_norm}
\end{figure}

Missing the climate extremes induced abnormal TWSA signals in the data records may not only introduce significant uncertainties and biases in the hydrological models during the data assimilation processes~\cite{AlexSun2020}, but also decrease the reliability of the GRACE-based drought/flood characterizations and assessments~\cite{AghaKouchak2015,Thomas2014}. Thus, we further analyze BCNN's performance in detecting dry and wet events during the gap (July 2017-May 2018). To this end, the trend and seasonal signals are removed from the original BCNN TWSAs, which is done by fitting a linear trend via unweighted least squares, plus annual and semiannual sinusoids to the TWSA time series~\cite{CHEN2014130,Zhong_2018}. The detrended and deseasonalized TWSAs at each grid point is then standardized using the $z$-score formula. The resulting TWSA is denoted as sTWSA$_{\mathrm{BCNN}}^{\mathrm{detrend,deseason}}$ and its spatial maps in the eleven missing months are shown in Figure~\ref{fig:TWS_intersubseas_norm}(a-k). It is observed that many regions exhibit abnormal signals during the gap, with the extremely negative and positive signals indicating dry and wet conditions, respectively. Here we focus on the four regions labeled with (I-IV) in Figure~\ref{fig:TWS_intersubseas_norm}h, as they have negative (regions I-III) or positive (region IV) sTWSA$_{\mathrm{BCNN}}^{\mathrm{detrend,deseason}}$ signals in several consecutive months during the gap. The dry/wet events in region I and regions III/IV  were also reported by the National Integrated Drought Information System (\url{https://www.drought.gov/states/arizona}) and the European State of the Climate (\url{https://climate.copernicus.eu/ESOTC}), respectively.

\begin{figure}[h!]
    \centering
    \includegraphics[width=\textwidth]{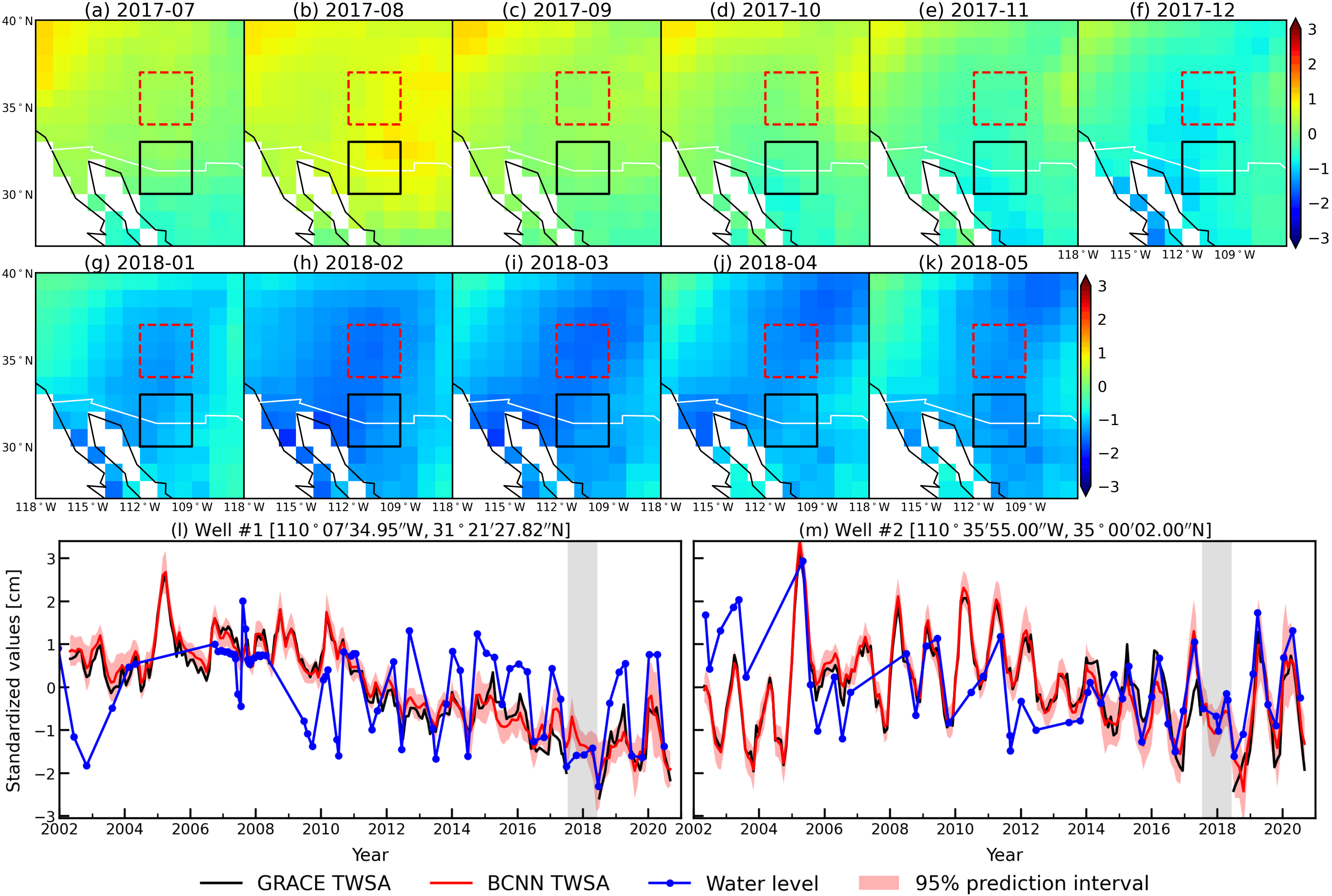}
    \caption{(a-k) The standardized signals of the detrended and deseasonalized BCNN TWSAs during the gap (July 2017-May 2018) in region I labeled in Figure~\ref{fig:TWS_intersubseas_norm}h. (l) and (m) illustrate the time series of groundwater levels, and the region-averaged original GRACE and BCNN TWSAs in regions around the observation well \#1 (black rectangle) and well \#2 (red dashed rectangle), respectively. The time series have been respectively standardized to facilitate the comparison and the  shaded  area  denotes the gap period, which experienced a dry event.}
    \label{fig:TWS_intersubseas_region_swUS}
\end{figure}

The four detected dry/wet events are examined using the time series of precipitation anomalies, standardized precipitation-evapotranspiration index (SPEI), and ground/surface water levels. Figures~\ref{fig:TWS_intersubseas_norm}(l-o) shows the region-averaged times series of sTWSA$_{\mathrm{GRACE}}^{\mathrm{detrend,deseason}}$, sTWSA$_{\mathrm{BCNN}}^{\mathrm{detrend,deseason}}$, standardized precipitation anomalies (s$P_{\mathrm{anom}}$) and 6-month SPEI. When calculating $P_{\mathrm{anom}}$, we first applied moving average on the precipitation time series ($P_t=\sum_{i=-3}^0P_{t+i}$) to smooth out short-term fluctuations, and then compute the anomalies for each month with respect to its respective average during 2002 and 2018. The SPEI dataset, which covers a period from 2002 to 2018, is downloaded from \url{https://spei.csic.es/spei_database/}. It is observed that the two sTWSA$^{\mathrm{detrend,deseason}}$ lines agree well with the s$P_{\mathrm{anom}}$ and SPEI lines over the period 2002-2018. This demonstrates the capability of sTWSA$^{\mathrm{detrend,deseason}}$ in detecting extreme climate events. More notably, the sTWSA$_{\mathrm{BCNN}}^{\mathrm{detrend,deseason}}$ signals in the four regions match well with the s$P_{\mathrm{anom}}$ and SPEI signals during the gap, where they exhibit consistently negative/positive extremes, suggesting the existence of the BCNN-identified dry/wet events.

\begin{figure}[h!]
    \centering
    \includegraphics[width=\textwidth]{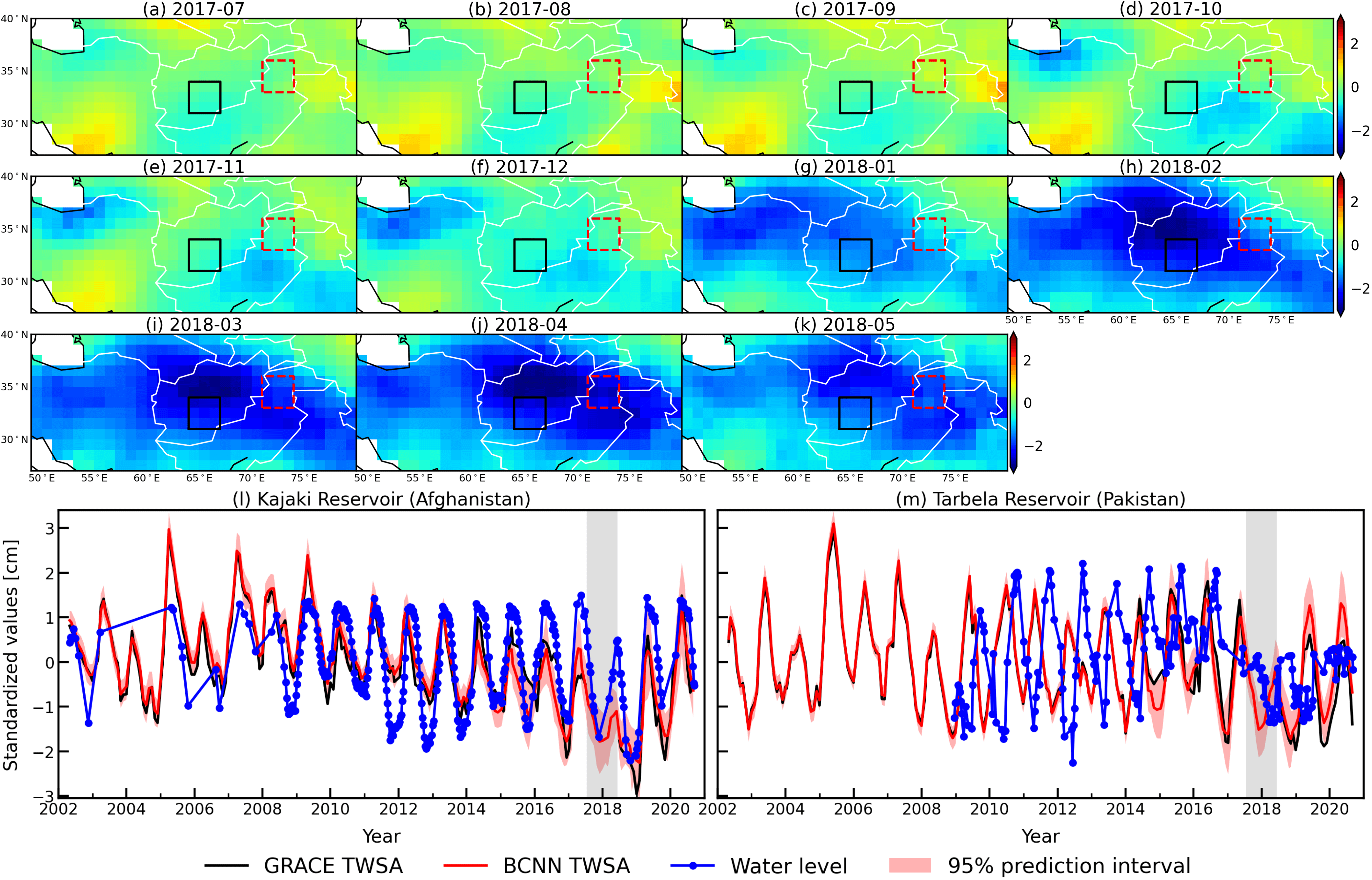}
    \caption{(a-k) The standardized signals of the detrended and deseasonalized BCNN TWSAs during the gap (July 2017-May 2018) in region II labeled in Figure~\ref{fig:TWS_intersubseas_norm}h. (l) and (m) illustrate the time series of reservoir water levels, and the region-averaged original GRACE and BCNN TWSAs in regions around the Kajaki Reservoir (black rectangle) and Tarbela Reservoir (red dashed rectangle), respectively. The time series have been respectively standardized to facilitate the comparison and the  shaded area denotes the gap period, which experienced a dry event.}
    \label{fig:TWS_intersubseas_region_iran}
\end{figure}

\begin{figure}[h!]
    \centering
    \includegraphics[width=\textwidth]{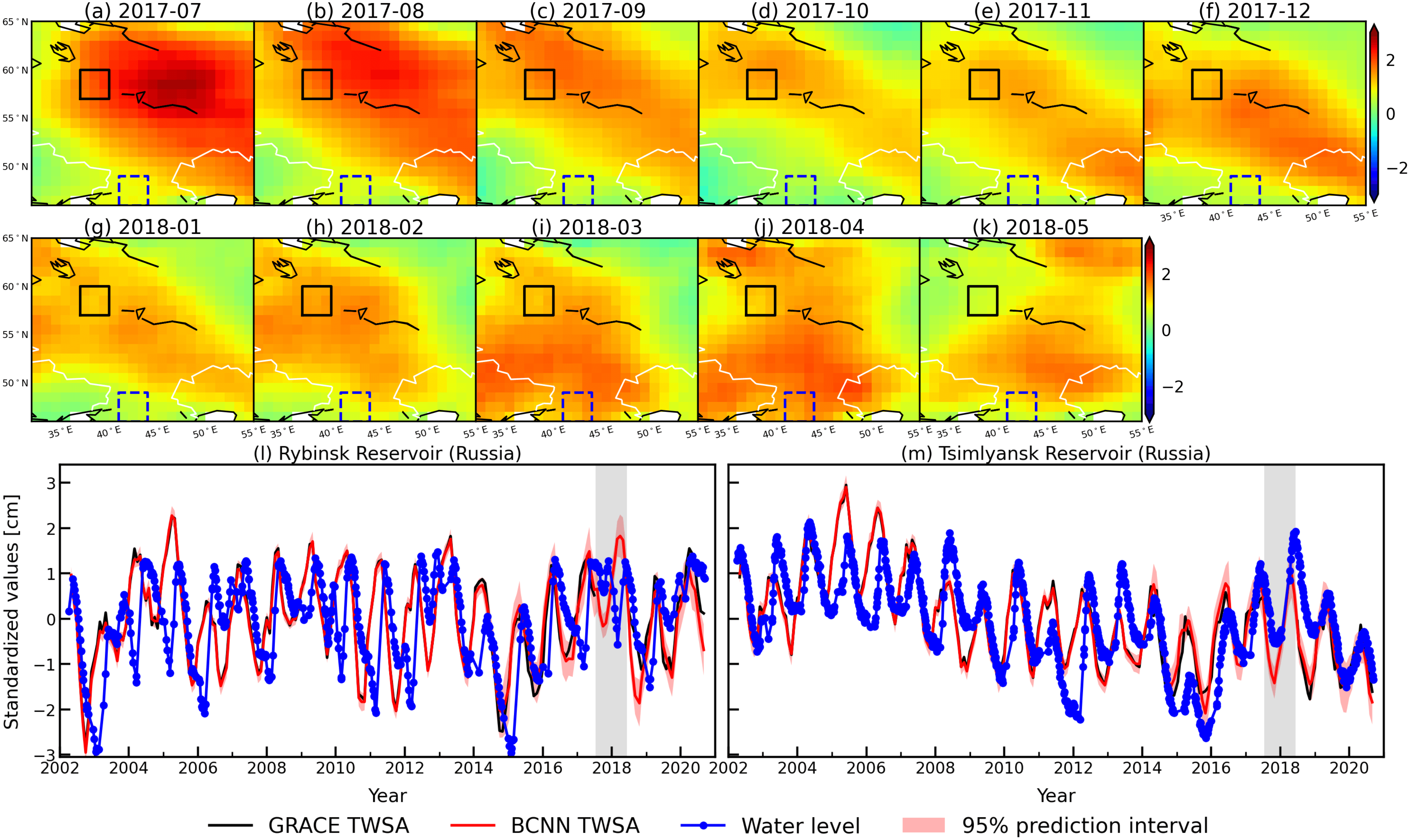}
    \caption{(a-k) The standardized signals of the detrended and deseasonalized BCNN TWSAs during the gap (July 2017-May 2018) in region II labeled in Figure~\ref{fig:TWS_intersubseas_norm}h. (l) and (m) illustrate the time series of reservoir water levels, and the region-averaged original GRACE and BCNN TWSAs in regions around the Rybinsk Reservoir (black rectangle) and Tsimlyansk Reservoir (blue dashed rectangle), respectively. The time series have been respectively standardized to facilitate the comparison and the  shaded area denotes the gap period, which experienced a wet event.}
    \label{fig:TWS_intersubseas_region_russia}
\end{figure}

The time series of ground/surface water levels, and GRACE and BCNN TWSAs at the observation locations within regions I, II, and IV are depicted in Figures~\ref{fig:TWS_intersubseas_region_swUS}-\ref{fig:TWS_intersubseas_region_russia}. The time series data have been respectively standardized to facilitate the comparison. Together, the sTWSA$_{\mathrm{BCNN}}^{\mathrm{detrend,deseason}}$ maps in the three regions during the 11-month gap are also shown. The groundwater level observations at two wells in region I were retrieved from the United States Geological Survey (\url{https://nwis.waterdata.usgs.gov/az/nwis/gwlevels}). The surface water levels of four reservoirs in regions III and IV measured by altimetry satellites are available at the Database for Hydrological Time Series of Inland Waters (\url{https://dahiti.dgfi.tum.de/}). The GRACE and BCNN TWSA time series are region-averaged of the nine grid cells around the observation locations (the rectangled regions in the plots). While there may exit phase deviations in the signals due to delayed responses or reservoir operations, it can be seen that the time series of ground/surface water level changes show a good agreement with the GRACE and BCNN TWSAs. The declined groundwater levels in region I (Figure~\ref{fig:TWS_intersubseas_region_swUS}) and reservoir water levels in region II (Figure~\ref{fig:TWS_intersubseas_region_iran}) during the gap indicate the occurrence of droughts in the two regions, while the increased reservoir water levels in region IV during the gap reflects the wet event (Figure~\ref{fig:TWS_intersubseas_region_russia}). The successful recovery of these extreme events enables us, for instance, to characterize and quantify the drought in the region around the Kajaki Reservoir (the water storage started to decrease in early 2017 and the water loss lasted until late 2018 as shown in Figure~\ref{fig:TWS_intersubseas_region_iran}l) and the wet event in the region around the Rybinsk Reservoir (it experienced a relatively wet 2017/2018 winter (\url{https://climate.copernicus.eu/ESOTC}) and the water gain reached to the peak in April 2018 as indicated by Figure~\ref{fig:TWS_intersubseas_region_russia}l).

\begin{table}[h!]
\caption{Summary of the Experimental Settings in Previous Studies.}
\centering
\label{tab:exp_set}
\begin{threeparttable}
\begin{tabularx}{\textwidth}{ 
  >{\raggedright\arraybackslash}X 
  >{\centering\arraybackslash}X 
  >{\raggedright\arraybackslash}X 
  >{\raggedright\arraybackslash}X 
  >{\raggedright\arraybackslash}X 
  >{\raggedright\arraybackslash}X}
\hline
Authors                                    & Study area            & GRACE data$^{\rm{a}}$       & Training period         & Test period   
\\
\hline
\citeA{Humphrey-2019-REC}$^{\rm{b}}$ & Global                & JPL mascon RL06 & \multicolumn{1}{c}{-}                       & Apr. 2014- June 2017; June 2018- July 2019 \\
\citeA{LietalWRR2020}$^{\rm{c}}$     & 26 basins                & CSR mascon RL06$^{\rm{d}}$ & Apr. 2002- June 2017    & June 2018- Dec. 2018                   \\
\citeA{SunZL2020WRR}$^{\rm{e}}$      & Global & CSR mascon RL06 & Apr. 2002- Jan. 2014 & Feb. 2014- June 2017 \\  
\hline
\end{tabularx}
\begin{tablenotes}
      \small
      \item $^{\rm{a}}$There might be multiple GRACE data products used in the referenced studies. Here we list the products used in the comparison.
      \item $^{\rm{b}}$The generated data product is known as GRACE-REC and available at \url{https://doi.org/10.6084/m9.figshare.7670849}. The product driven by the ERA5 climatic data is used for comparison.
      \item $^{\rm{c}}$The generated data product is available at \url{https://github.com/strawpants/twsc_recon}.
      \item $^{\rm{d}}$The CSR (Center for Space Research) GRACE product is downloaded from \url{http://www2.csr.utexas.edu/grace/}.
      \item $^{\rm{e}}$The generated data product is obtained upon request from the authors.
    \end{tablenotes}
  \end{threeparttable}
\end{table}

In summary, the sTWSA$_{\mathrm{BCNN}}^{\mathrm{detrend,deseason}}$ maps suggest that there were dry events in regions (I-III) and wet events in region IV during the GRACE and GRACE-FO gap, which are demonstrated by the s$P_{\mathrm{anom}}$, SPEI, ground/surface water level signals. These results, together with those presented in section~\ref{sec:accuracy_assess}, indicate that BCNN can successfully captures the complex spatio-temporal behaviors of GRACE TWSAs as well as the abnormal signals caused by climate extremes. It is worth noting that the test data from GRACE (April 2014-June 2017) and GRACE-FO (June 2018-Aug 2020) cover the periods before and after the gap (i.e. July 2017-May 2018). The good agreement between BCNN and GRACE TWSAs suggests the reliability of BCNN in bridging the GRACE and GRACE-FO gap at a global scale.

\subsection{Comparison with Previous Studies}
As presented in section~\ref{sec:introduction}, there have been many efforts undertaken to model the GRACE TWSAs using data-driven methods. Here we restrict the comparison with~\citeA{Humphrey-2019-REC}, \citeA{LietalWRR2020}, and \citeA{SunZL2020WRR}, who provided publicly accessible global-scale TWSA prediction products. Note that the predicted TWSA product released by~\citeA{Humphrey-2019-REC} is known as GRACE-REC. The original GRACE-REC dataset provides the detrended and deseasonalized TWSAs. The trend and seasonal signals obtained from the GRACE TWSAs and~\citeA{Humphrey2017GRL}, respectively, have been added to the original GRACE-REC TWSAs for consistency. For a fair comparison, the GRACE TWSA data and the training periods used for BCNN network training are respectively the same as those employed in the three studies. The detailed descriptions of the three TWSA prediction products are summarized in Table~\ref{tab:exp_set}. 

\begin{figure}[h!]
    \centering
    \noindent\includegraphics[width=\textwidth]{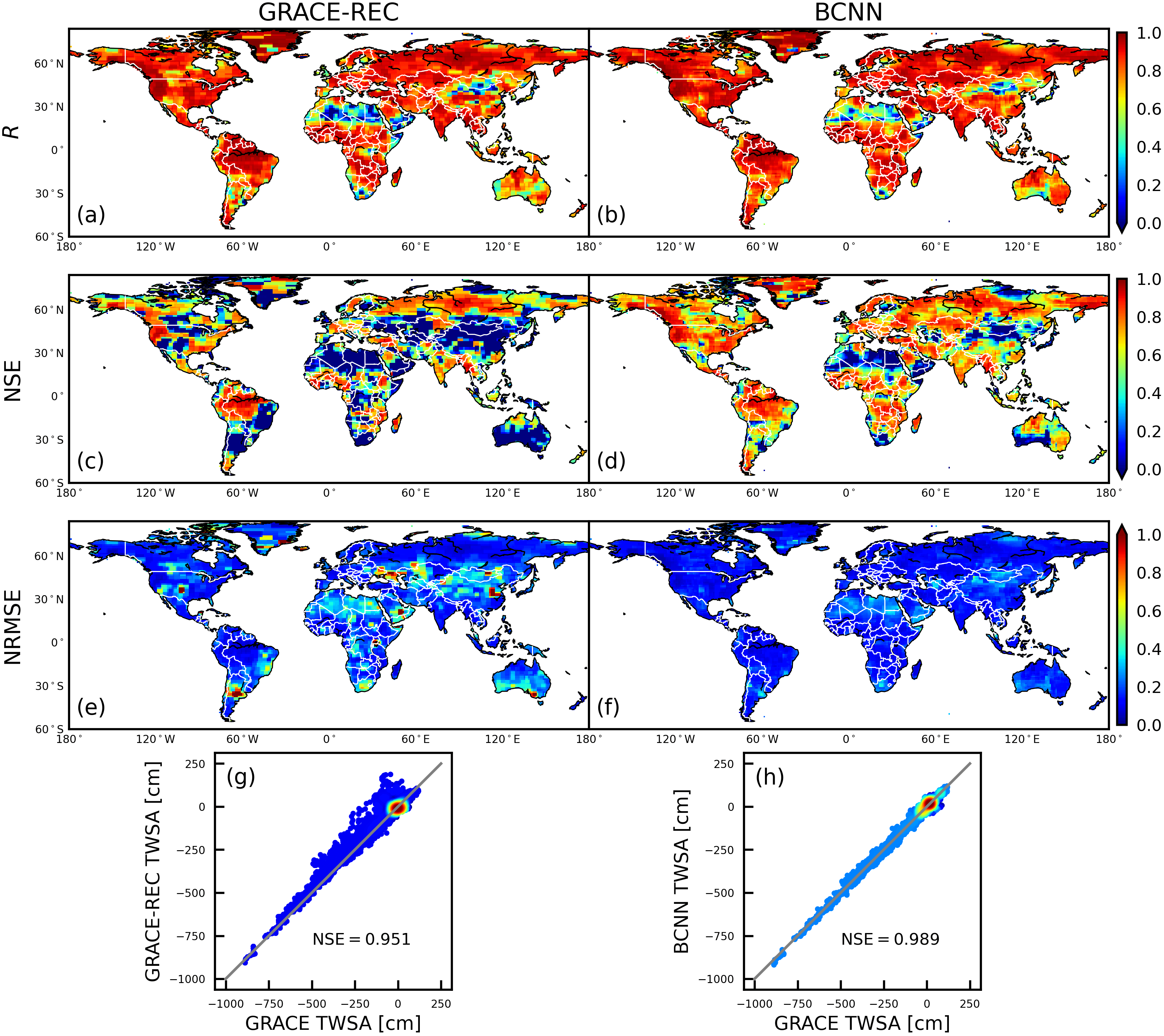}
    \caption{The predicted $R$ (row 1), NSE (row 2), and NRMSE (row 2) accuracy of GRACE-REC (left) and our BCNN (right) for the JPL mascon GRACE TWSAs (test period: April 2014-June 2017, June 2018-July 2019). Row 4: The density scatter plots between the GRACE and modeled TWSAs. The trend and seasonal signals extracted from GRACE TWSAs and~\citeA{Humphrey2017GRL}, respectively, have been added to the original GRACE-REC TWSAs for consistency.}
    \label{fig:R_RMSE_BCNNvsREC}
\end{figure}

\begin{figure}[h!]
    \centering
    \noindent\includegraphics[width=\textwidth]{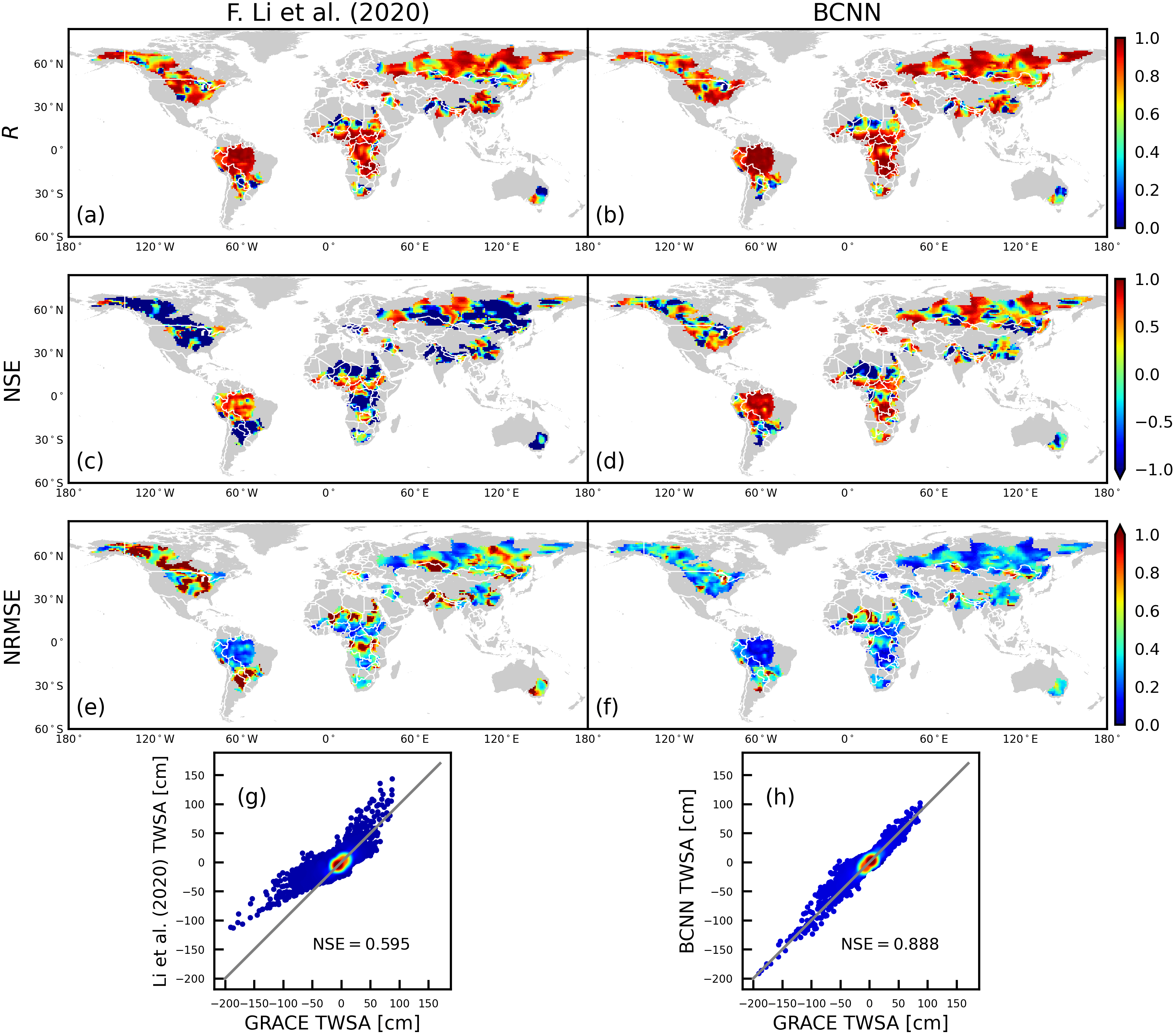}
    \caption{The predicted $R$ (row 1), NSE (row 2), and NRMSE (row 3) accuracy for the CSR mascon GRACE TWSAs obtained  by~\citeA{LietalWRR2020} (left) and our BCNN (right). Row 4: The density scatter plots between the GRACE and modeled TWSAs. The test period is June 2018-December 2018. }
    \label{fig:R_RMSE_BCNNvsLiFP}
\end{figure}

\begin{figure}[h!]
    \centering
    \noindent\includegraphics[width=\textwidth]{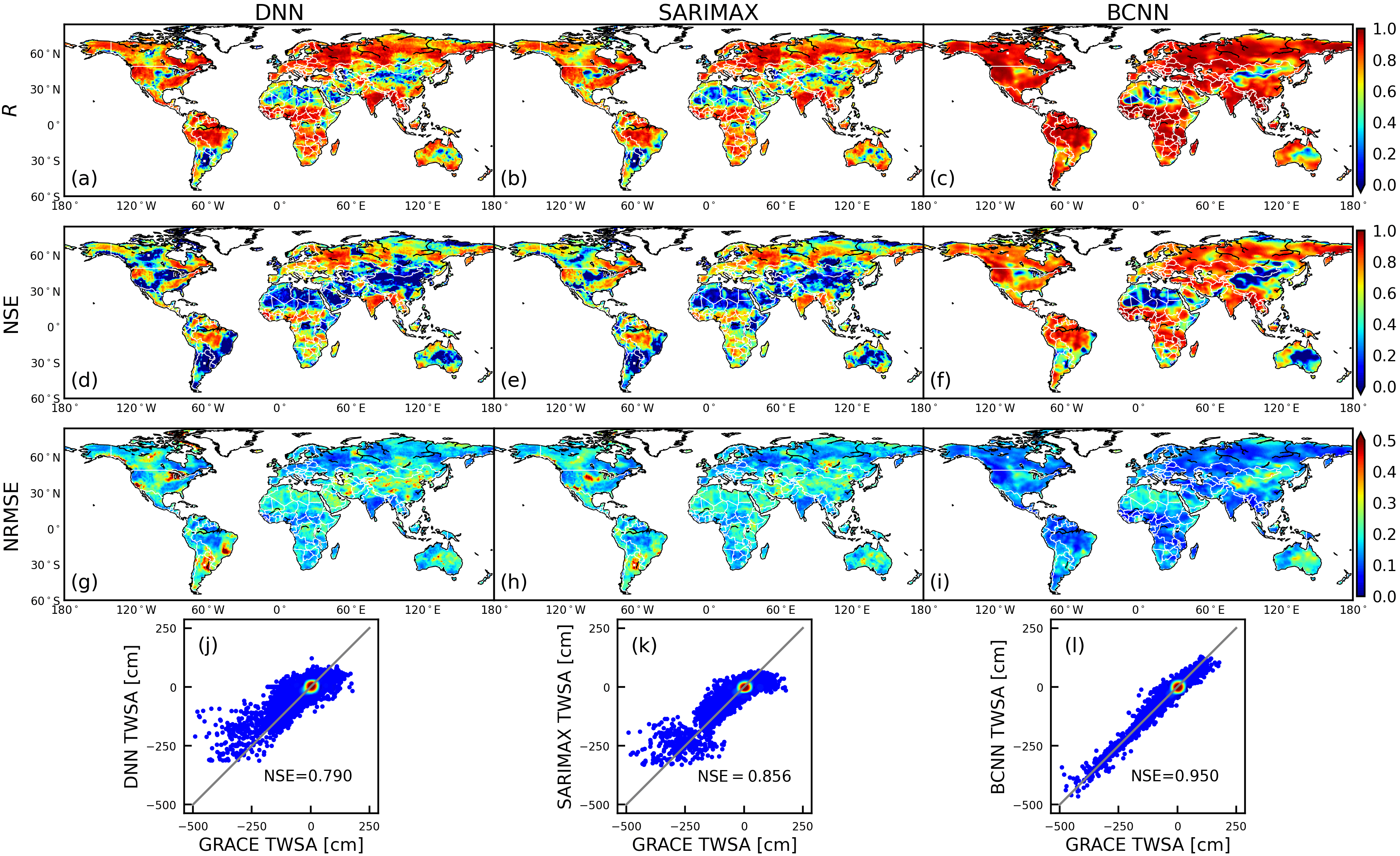}
    \caption{The predictive $R$ (row 1), NSE (row 2), and NRMSE (row 3) accuracy for the CSR mascon GRACE TWSAs obtained by the deep neural network (DNN) and seasonal autoregressive integrated moving average with exogenous variables (SARIMAX) methods employed in~\citeA{SunZL2020WRR} and our BCNN. The test period is February 2014-June 2017. }
    \label{fig:R_NSE_RMSE_BCNNvsSunZL}
\end{figure}

\begin{figure}[h!]
    \centering
    \noindent\includegraphics[width=\textwidth]{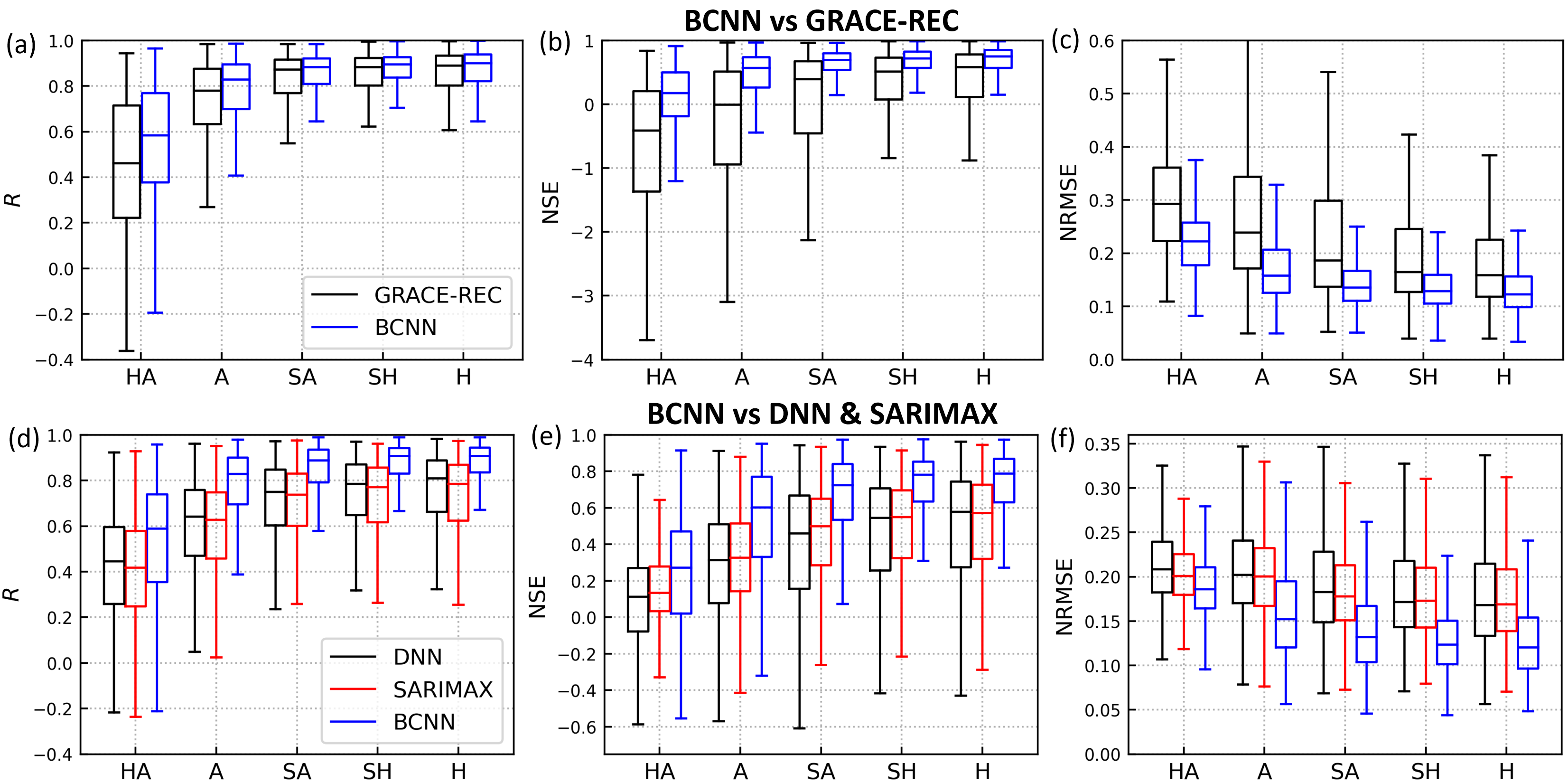}
    \caption{Boxplots of the $R$, NSE, and NRMSE values at the grid cells in the hyper-arid (HA), arid (A), semi-arid (SA), semi-humid (SH), and humid (H) regions. The outliers are not shown. (a-c) present the comparison between BCNN and GRACE-REC. (d-f) present the comparison between BCNN and \citeA{SunZL2020WRR} (i.e., the DNN and SARIMAX methods). Note that the training/test periods in the comparison (Table~\ref{tab:exp_set}) are different from that in Figure~\ref{fig:boxplot_vsNoah_ERA5L} (section~\ref{sec:net_arc_train}).}
    \label{fig:boxplot_vsREC_SunZL}
\end{figure}

The comparison results are shown in Figures~\ref{fig:R_RMSE_BCNNvsREC}-\ref{fig:R_NSE_RMSE_BCNNvsSunZL}, which illustrate the $R$, NSE, and NRMSE maps and the density scatter plots between the predicted and GRACE TWSAs. In addition, we also compare separately the performances in the hyper-arid, arid, semi-arid, semi-humid, and humid regions in Figure~\ref{fig:boxplot_vsREC_SunZL}, which shows the boxplots of the grid-cell-scale accuracy metrics in the five classes of regions (\citeA{LietalWRR2020} provided predictions for only 26 river basins, thus their results are not summarized). The medians of the boxplots are summarized in Tables~\ref{tab:medians_BCNNvsREC} and~\ref{tab:medians_BCNNvsSunZL}. Figures~\ref{fig:R_RMSE_BCNNvsREC}-\ref{fig:boxplot_vsREC_SunZL} obviously indicate BCNN's better performance in providing higher prediction accuracy for GRACE TWSAs relative to the methods proposed in the three previous studies. The previous methods~\cite{Humphrey-2019-REC,LietalWRR2020,SunZL2020WRR} obtain relatively high accuracy in the humid/semi-humid regions but their performances decrease in the hyper-arid/arid/semi-arid regions. Our BCNN method improves the prediction accuracy in these hyper-arid/arid/semi-arid regions to a relatively high level. For instance, compared to \citeA{Humphrey-2019-REC}, whose median NSE  values in the hyper-arid, arid, and semi-arid regions are -0.411 -0.005, and 0.395, respectively, while those of our BCNN have been improved to 0.175, 0.571, and 0.694, respectively (Table~\ref{tab:medians_BCNNvsREC}). In comparison to \citeA{SunZL2020WRR}, the median NSE  values in the three regions have been increased from 0.133, 0.326, and 0.500 to 0.272, 0.601, and 0.725 (Table~\ref{tab:medians_BCNNvsSunZL}). For our BCNN method, the median NSE values in the arid/semi-arid/semi-humid/humid are over a relatively high value of 0.55 (Tables~\ref{tab:medians_BCNNvsREC} and~\ref{tab:medians_BCNNvsSunZL}). While BCNN obtains a better accuracy in the hyper-arid region compared to previous studies, the still relatively low median NSE values (0.175 and 0.272) in this region indicate that its performance deserves further improvement. The insufficient GRACE signal-to-noise ratios (i.e. low signal variability) in the hyper-arid region may be the major obstacle to accuracy improvement \cite{SunZL2020WRR}.  

Recall that the GRACE trend information has been added to the original GRACE-REC TWSAs and was also utilized in~\citeA{LietalWRR2020} when predicting TWSAs. The results suggest the superior capability of BCNN in feature mining and thus providing improved TWSA predictions to bridge the GRACE and GRACE-FO gap. Two additional noteworthy advantages of BCNN over existing methods are the relatively few assumptions and preprocessing involved and its ability to handle simultaneously the global scale.

\section{Conclusions}\label{sec:conclusions}
The GRACE/GRACE-FO TWSA observations, together with the hydrological models, have been vital tools for water-related studies at regional or even global scales. However, the approximately one-year gap of TWSA observations between the two GRACE missions may introduce significant biases and uncertainties in the models and consequently produces misleading predictions. In this study, we propose a deep learning-based BCNN method driven by climatic data to fill this gap. By leveraging and implementing recent advances of deep learning (e.g., the residual and dense connections, attention mechanisms, and Bayesian training strategy), the BCNN model is able to effectively and efficiently extract important information for TWSA predictions from multi-source input data. Results show that BCNN can successfully capture the complex spatio-temporal behaviors of TWSAs and identify the extreme dry/wet events during the gap. The comparisons with hydrological model outputs and previous studies further suggest that BCNN obtains a state-of-the-art performance in bridging the gap at a global scale. In particular, BCNN exhibits a clearly better performance in the relatively arid regions. The improvements in predicting the missing TWSA signals, especially those anomalies induced by climate extremes, can be of great significance for enhancing the reliability of the hydrological model predictions and our characterizations and assessments of the extreme climate events during the gap.

The outperformance of BCNN is mainly attributed to the use of TWSA trends, which are derived from the available GRACE/GRACE-FO data before and after the gap, and its outstanding performance in feature extraction. The long-term TWSA trends are induced by anthropogenic and/or natural factors and usually challenging-to-learn~\cite{Humphrey-2019-REC,LietalWRR2020,SunZL2020WRR}. The utilization of this trend information makes full use of the existing data and essentially eases the learning task for BCNN. The BCNN's robust capability in extracting key features for TWSA predictions is especially illustrated by the comparison with the TWSA prediction products in which the GRACE trend information was also employed. Note that we are concerned with bridging the gap between GRACE and GRACE-FO in the current work. For the TWSA reconstruction task, which aims to reconstruct the TWSAs for pre-2002 and is beyond the scope of this study, the trend information is unavailable. The performance of BCNN for such a task remains to be investigated.

\acknowledgments
The JPL GRACE Mascon data used in this study are available at \url{https://podaac.jpl.nasa.gov/dataset/TELLUS_GRAC-GRFO_MASCON_CRI_GRID_RL06_V2}; ERA5-Land data are available at \url{https://cds.climate.copernicus.eu}; Noah TWSA dataset is downloaded from \url{https://disc.gsfc.nasa.gov/}. This work was funded by the National Key Research and Development Program of China (2018YFC1800600), National Natural Science Foundation of China (41730856, 41874095, 41977157, 42002248, 42004073), China Postdoctoral Science Foundation (2020M681550), Jiangsu Planned Projects for Postdoctoral Research Funds (2020Z133), and Fundamental Research Funds for the Central Universities (020614380106). The authors acknowledge Dr. Yinhao Zhu from Qualcomm AI Research for his valuable suggestions on implementing the Bayesian training strategy. The BCNN codes and the predicted TWSA dataset generated in this work will be made available at \url{https://github.com/njujinchun/bcnn4grace} upon publication of this manuscript.

\appendix
\section{This section provides some tables and figures which support the discussion of this article}

\begin{table}[h!]
\caption{Medians of the $R$, NSE, and NRMSE values at the grid cells in the hyper-arid (HA), arid (A), semi-arid (SA), semi-humid (SH), and humid (H) regions obtained by Noah, ERA5L, corrected Noah (cNoah), corrected ERA5L (cERA5L), and our BCNN. cNoah and cERA5L TWSAs are obtained by correcting their respective original TWSAs with GRACE TWSAs' trend (equation~(\ref{eq:cNoah&cERA5L})). The JPL mascon GRACE data are used and the test periods are April 2014-June 2017 and June 2018-August 2020. Bold value indicates the best performance.}
\centering
\label{tab:medians_BCNNvsNoah&ERA5L}
\footnotesize
\resizebox{\textwidth}{15mm}{
\begin{tabular}{llcccccccccccccc}
\hline
       & \multicolumn{5}{l}{$R$}                                                                       & \multicolumn{5}{l}{NSE}                                                                                                                    & \multicolumn{5}{l}{NRMSE}                                                          \\ \cline{2-16}
       & HA                        & A              & SA             & SH             & H              & HA                             & A                              & SA                             & SH                             & H                              & HA             & A              & SA             & SH             & H              \\ \hline
Noah   & 0.251                     & 0.631          & 0.671          & 0.705          & 0.662          & -10.501                        & -3.232                         & -1.084                         & -1.019                         & -0.815                         & 0.758          & 0.479          & 0.346          & 0.340          & 0.318          \\
ERA5L  & 0.217                     & 0.605          & 0.745          & 0.759          & 0.742          & -8.509 & -2.005 & -0.728 & -0.564 & -0.263 & 0.692          & 0.405          & 0.313          & 0.295          & 0.268          \\
cNoah  & \multicolumn{1}{l}{0.493} & 0.678          & 0.686          & 0.729          & 0.734          & -0.411                         & 0.061                          & 0.183                          & 0.231                          & 0.238                          & 0.267          & 0.229          & 0.216          & 0.207          & 0.210          \\
cERA5L & \multicolumn{1}{l}{0.525} & 0.676          & 0.761          & 0.787          & 0.802          & 0.013                          & 0.284                          & 0.352                          & 0.393                          & 0.433                          & 0.221          & 0.201          & 0.192          & 0.185          & 0.181          \\
BCNN   & \textbf{0.639}            & \textbf{0.847} & \textbf{0.887} & \textbf{0.895} & \textbf{0.906} & \textbf{0.273}                 & \textbf{0.633}                 & \textbf{0.726}                 & \textbf{0.751}                 & \textbf{0.777}                 & \textbf{0.185} & \textbf{0.143} & \textbf{0.124} & \textbf{0.119} & \textbf{0.114} \\ \hline
\end{tabular}
}
\end{table}

\begin{table}[h!]
\caption{Medians of the $R$, NSE, and NRMSE values at the grid cells in the hyper-arid (HA), arid (A), semi-arid (SA), semi-humid (SH), and humid (H) regions obtained by GRACE-REC \cite{Humphrey-2019-REC} and our BCNN. The JPL mascon GRACE data are used and the test period are April 2014-June 2017 and June 2018-July 2019. Bold value indicates the best performance.}
\centering
\label{tab:medians_BCNNvsREC}
\footnotesize
\resizebox{\textwidth}{15mm}{
\begin{tabular}{llcccccccccccccc}
\hline
          & \multicolumn{5}{l}{$R$}                                                                       & \multicolumn{5}{l}{NSE}                                                                                                                    & \multicolumn{5}{l}{NRMSE}                                                          \\ \cline{2-16} 
          & \multicolumn{1}{c}{HA} & A              & SA             & SH             & H              & HA             & A              & SA             & SH             & H              & HA             & A              & SA             & SH             & H              \\ \hline
GRACE-REC & 0.462                  & 0.780          & 0.872          & 0.882          & 0.889          & -0.411         & -0.005         & 0.395          & 0.514          & 0.582          & 0.292          & 0.238          & 0.186          & 0.165          & 0.158          \\
BCNN      & \textbf{0.584}         & \textbf{0.828} & \textbf{0.882} & \textbf{0.894} & \textbf{0.899} & \textbf{0.175} & \textbf{0.571} & \textbf{0.694} & \textbf{0.718} & \textbf{0.751} & \textbf{0.222} & \textbf{0.158} & \textbf{0.135} & \textbf{0.129} & \textbf{0.122} \\ \hline
\end{tabular}
}
\end{table}

\begin{table}[h!]
\caption{Medians of the $R$, NSE, and NRMSE values at the grid cells in the hyper-arid (HA), arid (A), semi-arid (SA), semi-humid (SH), and humid (H) regions obtained by the DNN and SARIMAX methods employed in \citeA{SunZL2020WRR}, and our BCNN. The CSR mascon GRACE data are used and the test period is February 2014-June 2017. Bold value indicates the best performance.}
\centering
\label{tab:medians_BCNNvsSunZL}
\footnotesize
\resizebox{\textwidth}{15mm}{
\begin{tabular}{llcccccccccccccc}
\hline
        & \multicolumn{5}{l}{$R$}                                                                       & \multicolumn{5}{l}{NSE}                                                                                                                    & \multicolumn{5}{l}{NRMSE}                                                          \\ \cline{2-16}
        & \multicolumn{1}{c}{HA} & A              & SA             & SH             & H              & HA             & A              & SA             & SH             & H              & HA             & A              & SA             & SH             & H              \\ \hline
DNN     & 0.446                  & 0.641          & 0.750          & 0.784          & 0.809          & 0.113          & 0.314          & 0.460          & 0.545          & 0.578          & 0.209          & 0.202          & 0.183          & 0.171          & 0.168          \\
SARIMAX & 0.418                  & 0.628          & 0.737          & 0.770          & 0.785          & 0.133          & 0.326          & 0.500          & 0.550          & 0.571          & 0.201          & 0.200          & 0.178          & 0.173          & 0.169          \\
BCNN    & \textbf{0.590}         & \textbf{0.828} & \textbf{0.888} & \textbf{0.906} & \textbf{0.907} & \textbf{0.272} & \textbf{0.601} & \textbf{0.725} & \textbf{0.780} & \textbf{0.788} & \textbf{0.186} & \textbf{0.152} & \textbf{0.132} & \textbf{0.123} & \textbf{0.120} \\ \hline
\end{tabular}
}
\end{table}

\begin{figure}[h!]
    \centering
    \noindent\includegraphics[width=\textwidth]{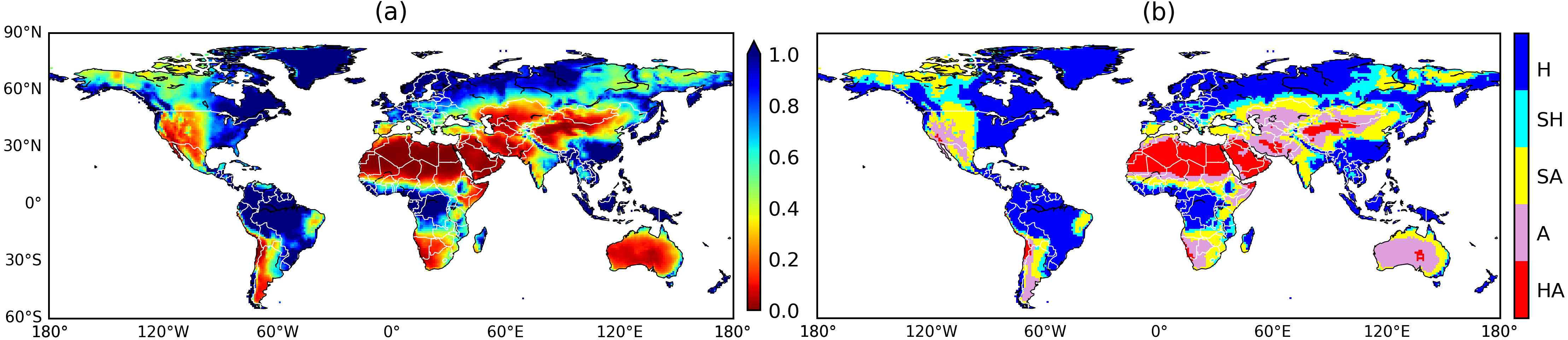}
    \caption{(a) A global map of the aridity index (AI), with lower and higher AI values indicating more arid and more humid conditions, respectively. (b) Spatial distribution of the hyper-arid (HA; AI$<$0.05), arid (A; 0.05$\leq$AI$<0.2$), semi-arid (SA; 0.2$\leq$AI$<0.5$), semi-humid (SH; 0.5$\leq$AI$<0.65$), and humid (H; AI$\geq$0.65) regions. The AI data are dowloaded from \url{https://doi.org/10.6084/m9.figshare.7504448.v3} \cite{Trabucco2019}.}
    \label{fig:AI}
\end{figure}

\begin{figure}[h!]
    \centering
    \noindent\includegraphics[width=\textwidth]{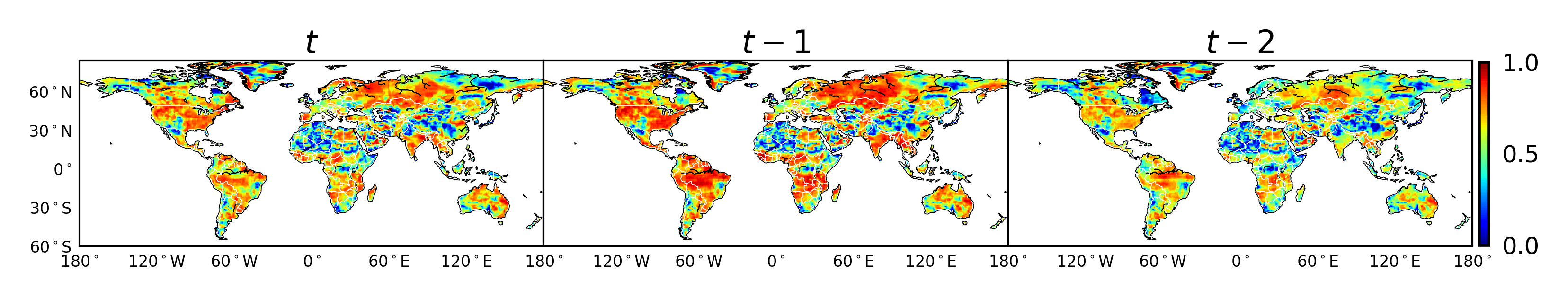}
    \caption{Spatial distribution of the absolute correlation coefficient ($|R|$) between the cumulative water storage changes (CWSCs) and GRACE TWSAs (April 2002-June 2017, June 2018-August 2020), where $(t-i)$ denotes $i$-month lag.}
    \label{fig:R_CWSCvsGRACE}
\end{figure}

\begin{figure}[h!]
    \centering
    \noindent\includegraphics[width=.8\textwidth]{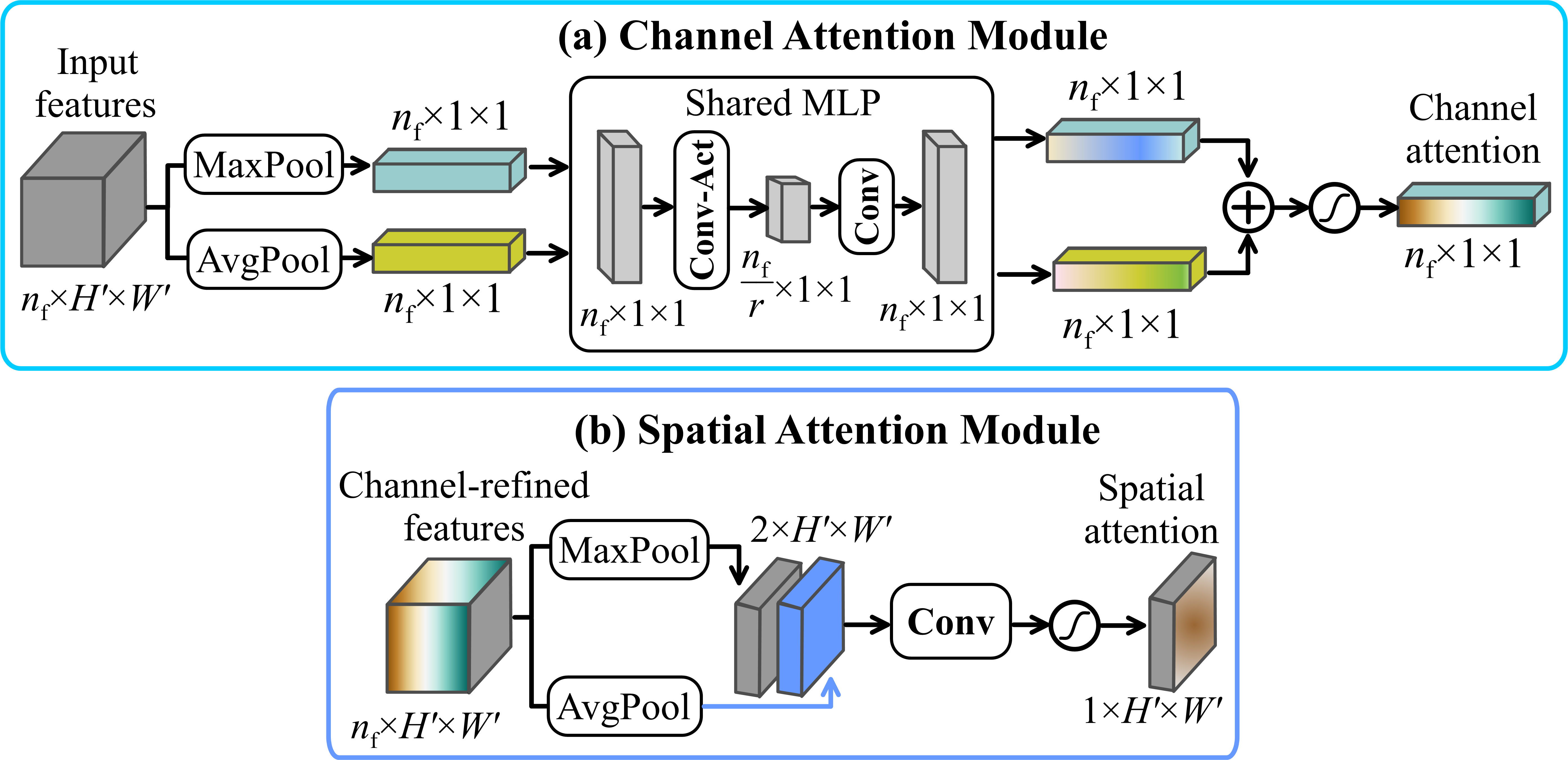}
    \caption{Diagrams of the channel and spatial attention modules. The inputs to each module are $n_{\rm{f}}$ feature maps with a size of $H^{'}\times W^{'}$. The channel attention module utilizes both max-pooling (MaxPool) outputs ($n_{\rm{f}}\times 1\times 1$) and average-pooling (AvgPool) outputs ($n_{\rm{f}}\times 1\times 1$) with a shared multi-layer perceptron (MLP) to produce a channel attention map. The spatial attention module utilizes similar two outputs (each with a shape of $1\times H^{'}\times W^{'}$) to produce a spatial attention map. The sigmoid activation is used to guarantee the values in the attention maps are between 0 and 1. Conv=Convolution; Act=Activation.}
    \label{fig:channel_spatial_attention}
\end{figure}

\begin{figure}[h!]
    \centering
    \noindent\includegraphics[width=\textwidth]{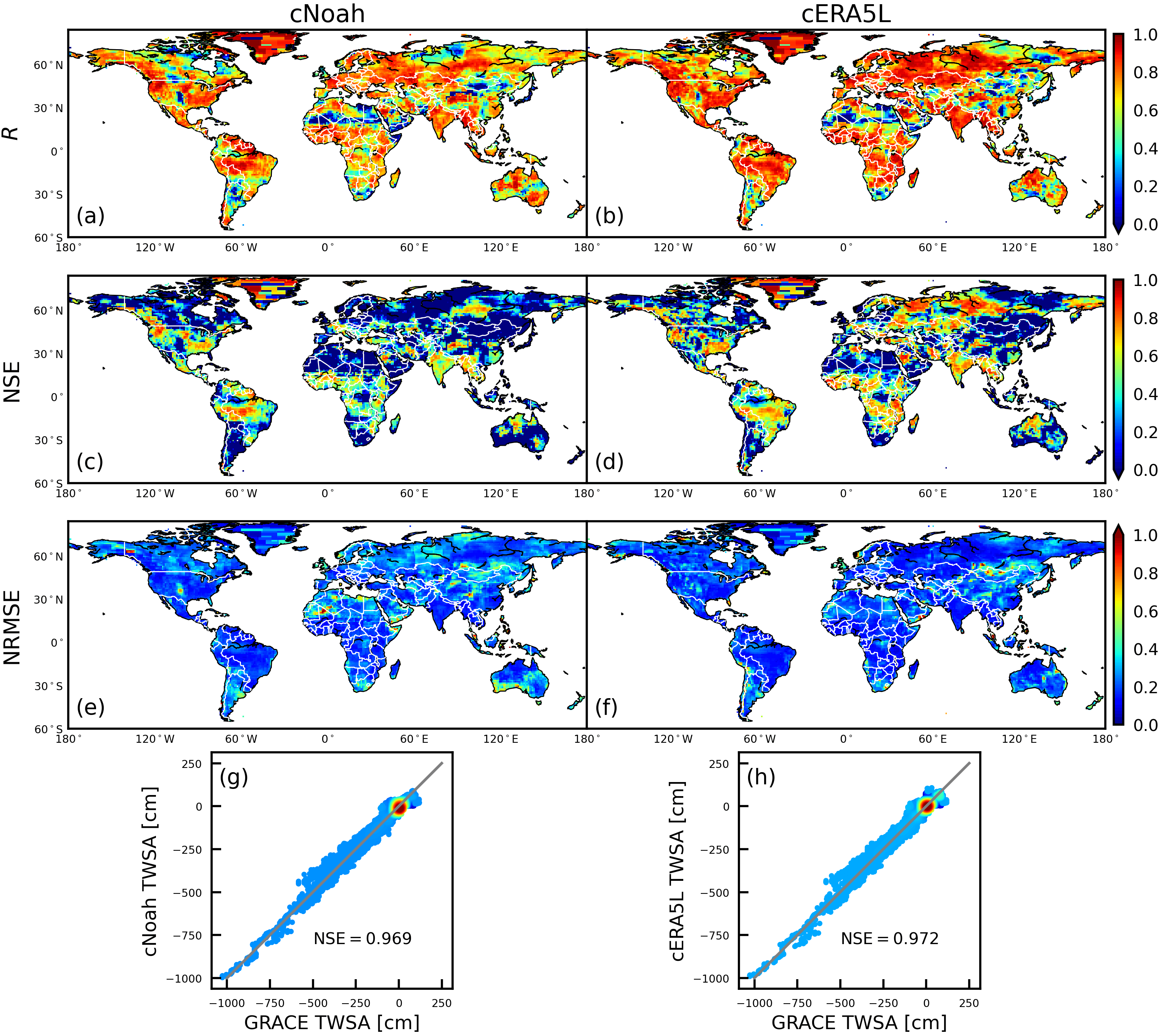}
    \caption{Spatial maps of $R$ (row 1), NSE (row 2), and NRMSE (row 3) between the GRACE TWSAs and the corrected Noah (cNoah)- (left) and corrected ERA5L (cERA5L)-derived (right) TWSAs during the test period (April 2014-June 2017, June 2018-August 2020). Row 4: The density scatter plots between the GRACE and modeled TWSAs. The cNoah and cERA5L TWSAs are obtained by correcting their respective original TWSAs with GRACE TWSAs' trend (equation~(\ref{eq:cNoah&cERA5L})).}
    \label{fig:metrics_Noah_ERA5L_trendGRACE}
\end{figure}

\begin{figure}[h!]
    \centering
    \noindent\includegraphics[width=\textwidth]{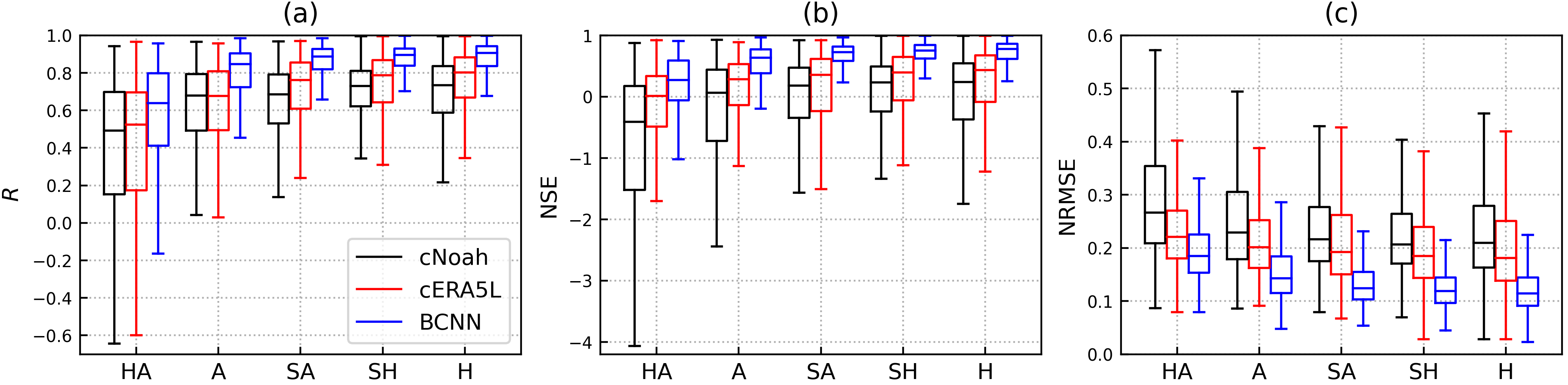}
    \caption{Boxplots of the (a) $R$, (b) NSE, and (c) NRMSE values at the grid cells in the hyper-arid (HA), arid (A), semi-arid (SA), semi-humid (SH), and humid (H) regions obtained by the corrected Noah (cNoah), corrected ERA5L (cERA5L), and our BCNN. The outliers are not shown. The test period are April 2014-June 2017 and June 2018-August 2020. The cNoah and cERA5L TWSAs are obtained by correcting their respective original TWSAs with GRACE TWSAs' trend (equation~(\ref{eq:cNoah&cERA5L})).}
    \label{fig:boxplot_vsNoah_ERA5L_GRACEtrend}
\end{figure}

\begin{figure}[h!]
    \centering
    \noindent\includegraphics[width=\textwidth]{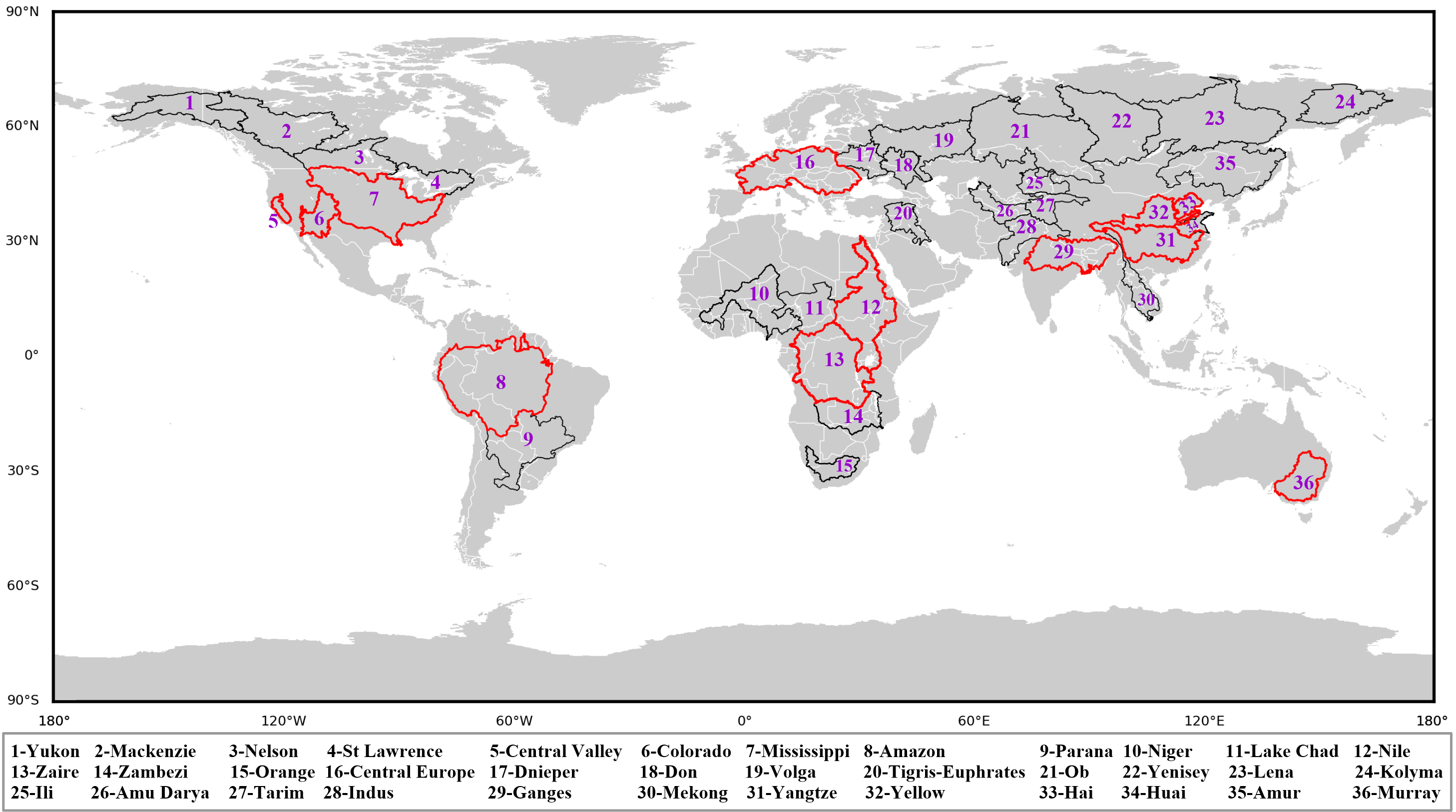}
    \caption{Locations of the regions and river basins considered for result analysis. Results of the regions/basins shown in red and black are shown in Figures~\ref{fig:pred_timeseries} and \ref{fig:time_series_SI}, respectively.}
    \label{fig:basin_map}
\end{figure}

\begin{figure}[h!]
    \centering
    \noindent\includegraphics[width=\textwidth]{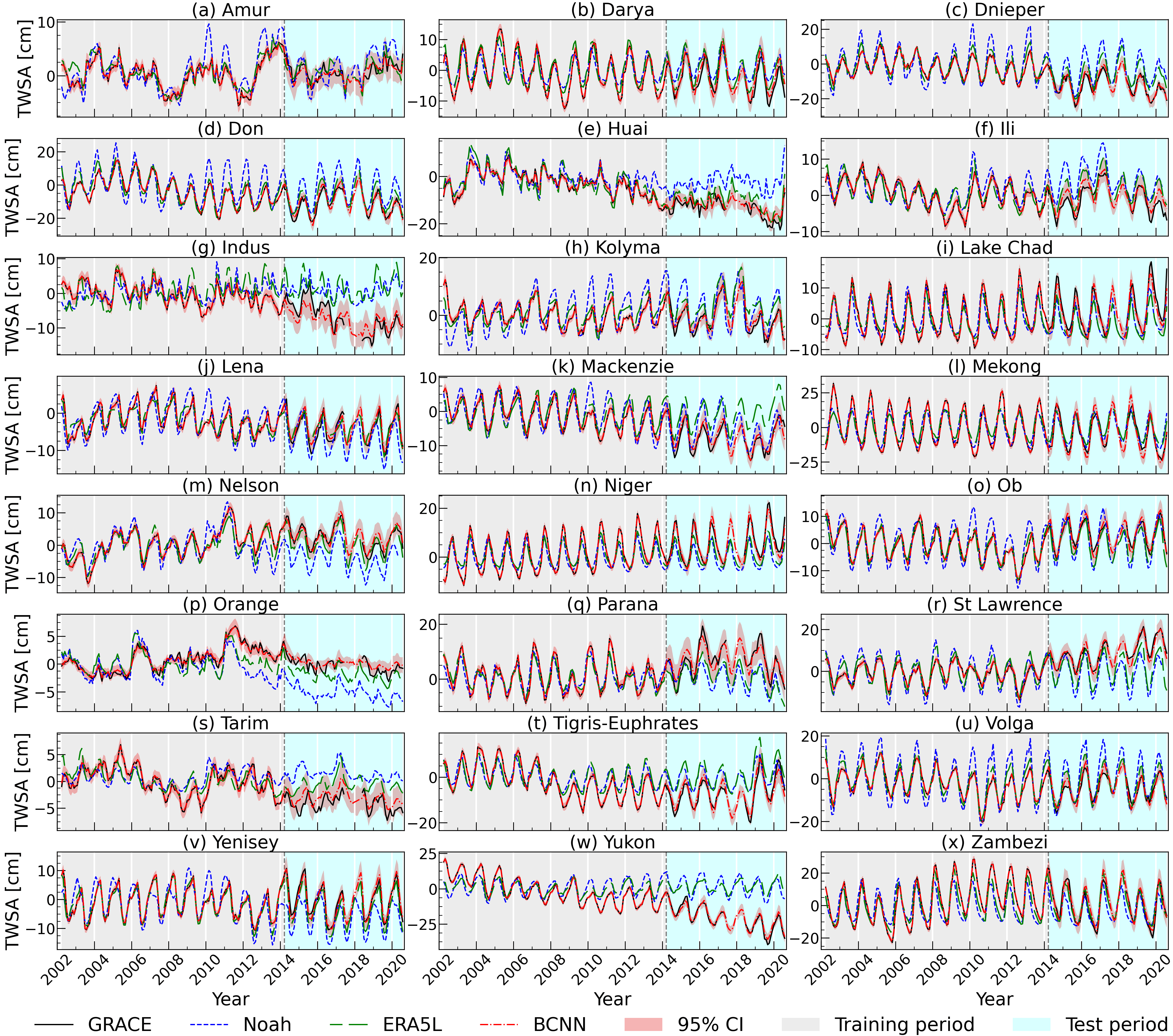}
    \caption{The basin-averaged GRACE, Noah, ERA5L, and BCNN TWSA time serials in different river basins. Shaded areas represent the 95\% confidence interval (CI) of BCNN predictions.}
    \label{fig:time_series_SI}
\end{figure}

\clearpage

\bibliography{main}

\end{document}